*Data Descriptor*

# A Large-Scale Dataset of Search Interests Related to Disease X Originating from Different Geographic Regions

Nirmalya Thakur [1,*], Shuqi Cui [1], Kesha A. Patel [2], Isabella Hall [3] and Yuvraj Nihal Duggal [1]

[1] Department of Computer Science, Emory University, Atlanta, GA 30322, USA; nicole.cui@emory.edu (S.C.); yuvraj.nihal.duggal@emory.edu (Y.N.D.)
[2] Department of Mathematics, Emory University, Atlanta, GA 30322, USA; kesha.patel@emory.edu
[3] Department of Computer Science, University of Cincinnati, Cincinnati, OH 45221, USA; hallib@mail.uc.edu
* Correspondence: nirmalya.thakur@emory.edu

**Abstract:** The World Health Organization (WHO) added Disease X to their shortlist of blueprint priority diseases to represent a hypothetical, unknown pathogen that could cause a future epidemic. During different virus outbreaks of the past, such as COVID-19, Influenza, Lyme Disease, and Zika virus, researchers from various disciplines utilized Google Trends to mine multimodal components of web behavior to study, investigate, and analyze the global awareness, preparedness, and response associated with these respective virus outbreaks. As the world prepares for Disease X, a dataset on web behavior related to Disease X would be crucial to contribute towards the timely advancement of research in this field. Furthermore, none of the prior works in this field have focused on the development of a dataset to compile relevant web behavior data, which would help to prepare for Disease X. To address these research challenges, this work presents a dataset of web behavior related to Disease X, which emerged from different geographic regions of the world, between February 2018 and August 2023. Specifically, this dataset presents the search interests related to Disease X from 94 geographic regions. These regions were chosen for data mining as these regions recorded significant search interests related to Disease X during this timeframe. The dataset was developed by collecting data using Google Trends. The relevant search interests for all these regions for each month in this time range are available in this dataset. This paper also discusses the compliance of this dataset with the FAIR principles of scientific data management. Finally, an analysis of this dataset is presented to uphold the applicability, relevance, and usefulness of this dataset for the investigation of different research questions in the interrelated fields of Big Data, Data Mining, Healthcare, Epidemiology, and Data Analysis with a specific focus on Disease X.

**Dataset:** https://dx.doi.org/10.21227/ht7f-rx42

**Dataset License:** CC BY 4.0

## 1. Introduction

In the recent past, several viruses, such as COVID-19 [1], the plague [2], the Spanish flu [3], HIV [4], Ebola [5], and MPox [6], have rampaged unopposed across different countries, infecting and leading to the demise of people, resulting in the destruction of political regimes, affecting various sectors of the global economy, and causing financial and psychosocial burdens, the likes of which the world has not witnessed in centuries [7]. As a response to this, various organizations and policy-making bodies on a global scale have begun investigating approaches to learn from such virus outbreaks, with an aim to not repeat the mistakes of the past during future virus outbreaks. "Disease X" is a placeholder name that was adopted by the World Health Organization (WHO) in February 2018 in their shortlist of blueprint priority diseases to represent a hypothetical, unknown pathogen that could cause a future epidemic [8,9]. The WHO used the placeholder term "Disease X" to make sure that its planning (such as relevant tests, expanded vaccinations, and production capabilities for vaccines) was robust, versatile, and equipped to deal with an unidentified virus [10]. The idea of Disease X, according to Anthony Fauci (the director of the US National Institute of Allergy and Infectious Diseases at that time), was to motivate the WHO's investigations on entire classes of viruses rather than just specific strains of certain viruses, with an aim to strengthen the WHO's preparedness for dealing with such outbreaks [11].

Thus, it is crucial to plan and adopt a holistic approach to prevent and predict a new pandemic in the future. Prior works [12–15] in this field have discussed various means by which Disease X might start. For instance, the potential of deadly pathogens being released from melting glaciers could start a new pandemic. Alternatively, with the continual increase of global warming and climate changes, viruses dormant at present may become active and mutate and lead to the next pandemic. Furthermore, human and animal contact has become increasingly common, and the lack of proper protocols in this regard has led to the outbreak of zoonotic viruses in the past. A well-known example of this would be H1N1, which contained genetic material from human, avian, and swine origin, involving wildlife, pig farming, animal movement, and farm workers [16]. Therefore, in the last couple of years or so, works in this field have also focused on predicting what type of pathogen might be responsible for Disease X, with an aim to create, implement, and evaluate countermeasures that would help control the potential pandemic at a faster rate than previous pandemics, such as COVID-19 [17–19]. Simpson et al. [20] stated that Disease X is likely to occur due to one or more of these risk factors—human interactions with wildlife, the production of goods derived from animals with minimal oversight of workers and an unclear supply chain, bug and tick vectors, extremely high population densities, and limited surveillance and laboratory capacities. This work by Simpson et al. [20] also states that Disease X will probably be caused by the zoonotic spread of a highly infectious RNA virus from a region where the confluence of risk factors and population dynamics will lead to prolonged person-to-person transmission.

A highly agreed upon aspect related to Disease X within the research community in this field is that the world is currently not prepared with the applicable countermeasures, policies, and procedures that would be necessary to control and contain this virus. There are multiple factors that we need to take into consideration when creating new response, control, and preparation measures, including vaccine development and distribution, country and state responses, political stances, and cultural and environmental factors. It is crucial that there is global preparation, coordination, and communication such that each of these factors is considered and managed in coordination with other factors to allow for controlling and containing a new pandemic [21].

One of the overarching issues that was observed on a global scale when attempting to handle the COVID-19 pandemic was the lack of efficiency, coordination, agreement, and organization related to the production and distribution of vaccines and COVID-19 tests in a timely manner [22]. While various organizations and labs were working to create a vaccine, it seemed that different countries were scrambling to even put-up testing centers and mass produce enough COVID-19 tests. It took far longer than ideal to ensure easy access to COVID-19 tests, which allowed the COVID-19 virus to continue to spread at an alarming rate because symptoms were not guaranteed to be noticeable in all population groups [23,24]. Testing is one of the first lines of defense against viruses because the threshold of the spread can be determined, and suitable actions can be taken depending on the positive cases that are reported [25]. This was an issue with the supply chain and communication across agencies during the outbreak and rapid spread of COVID-19. Those same supply chain issues were reported when trying to roll out the COVID-19 vaccines at an even slower rate than the tests. Research labs in different geographic regions seemed less prepared to mass produce and distribute the vaccines, which slowed down response rates and did not contain the spread of COVID-19 in a timely manner [26,27]. Another issue associated with the COVID-19 pandemic was the lack of coordination and cooperation between countries in their responses [28]. During the outbreak of COVID-19, some countries implemented measures (such as partial or complete lockdowns) immediately, while others did not implement such measures at the same pace [29,30]. Finally, a major issue specifically seen during the COVID-19 pandemic was political stances standing in the way of scientific progress. A lot of misinformation circulated over the course of the pandemic, which included the effectiveness of vaccines, the safety of the vaccines, the accuracy of the test results, approaches for treatment, and the severity of the virus. [31,32].

During the outbreak of COVID-19 and similar viruses in the past, Google Trends attracted a significant amount of attention from researchers across different disciplines, such as Big Data [33,34], Data Mining [35,36], Healthcare [37,38], Epidemiology [39,40], Information Retrieval [41,42], and Data Analysis [43,44], as Google Trends helps to mine, analyze, and obtain real-time insights related to web behavior, and the features of Google Trends surpass traditional surveys [45]. However, none of the prior works in this field have focused on mining and analyzing web behavior data from Google Trends related to Disease X. Addressing this research gap serves as the main motivation for this work.

The rest of this paper is organized as follows. In Section 2, a review of recent works in this field is presented. Section 3 discusses the methodology that was followed for mining relevant web behavior data from Google Trends to develop the proposed dataset. The description of the dataset is presented in Section 4. Section 4 also highlights how the dataset complies with the FAIR principles of scientific data management. A brief analysis of the dataset and potential applications of this dataset are presented in Section 5. The conclusion is presented in Section 6, which is followed by references.

## 2. Literature Review

In today's "Internet of Everything" style of living, the internet is integrated into our everyday routines more than ever before [46]. The web affords people across the world opportunities for internet-mediated engagement, and many of our everyday activities and lifestyles are rapidly transitioning to activities being conducted on the internet. Therefore, mining, analysis, and modeling of web behavior hold substantial importance across various disciplines, particularly for enhancing recommender systems [47], collaborative filtering mechanisms [48], user behavior clustering [49], customization of technology [50], modeling user trust and acceptance towards emerging technologies [51], enhancing webpage transitions [52], user personality detection [53], user interest analysis [54], monitoring virus outbreaks [55], and forecasting epidemics [56], just to name a few.

There have been multiple methodologies developed, implemented, and applied for web behavior mining, modeling, and analysis. However, in the last few years, Google Trends has become increasingly popular amongst researchers from different disciplines for such research works related to studying web behavior [57]. In the area of web behavior monitoring and analysis associated with different virus outbreaks, Google Trends has had a wide range of applications and use cases [58–61]. Ginsberg et al. [62] discussed the significance of seasonal influenza and the potential threat of a pandemic caused by a new strain of the influenza virus using Google Trends. The work proposed a method to enhance early disease detection by monitoring Google Search queries, which reflected health-seeking behavior. By analyzing Google Search queries, the researchers accurately estimated weekly cases of influenza in different regions of the United States, allowing for rapid detection and response to influenza with only a one-day reporting lag. The work by Kapiány-Fövény et al. [63] focused on analyzing Google Search volumes using Google Trends to forecast Lyme disease incidences. By integrating Google Trends data into a seasonal autoregressive moving average (SARIMA) model, the researchers compared their predictions with the actual reported values for Lyme disease incidence in Germany. The objective of the work done by Verma et al. [64] was to assess the potential of using Google

Trends data for predicting disease outbreaks. Focusing on diseases like malaria, dengue fever, chikungunya, and enteric fever in two regions in India—Chandigarh and Haryana— the research compared Google Search trends with Integrated Disease Surveillance Programme (IDSP) data. The analysis revealed a temporal correlation between the two datasets, particularly with a lag of 2 to 3 weeks for chikungunya and dengue fever, indicating the feasibility of utilizing Google Trends for predicting disease outbreaks at both local and regional levels. Young et al. [65] explored the potential of using relevant Google Search queries from Google Trends to monitor and predict syphilis cases at a state level. The study investigated the relationship between weekly reported syphilis cases and online search activity related to risk factors. By employing linear mixed models, the study established associations between search query data and syphilis cases, achieving accurate predictions for a significant number of weeks. The results indicated a strong correlation between web behavior and reported syphilis cases, suggesting the feasibility of integrating such data into public health monitoring systems for disease surveillance and prediction. Another work by Young et al. [66] focused on utilizing Google Search data to monitor and predict new HIV diagnosis cases in the United States. They collected HIV-related search volume data and state-level new HIV diagnoses data using Google Trends. Thereafter, they developed a predictive model using significant predictor keywords identified through LASSO and combined this data with actual HIV case reports from the CDC. The model demonstrated strong predictive capabilities, achieving an average $R^2$ value of 0.99 and an average root-mean-square error (RMSE) of 108.75 when comparing predicted and actual HIV cases. Morsy et al. [67] focused on predicting Zika virus cases using Google Search queries from Google Trends. The researchers developed a prediction model based on time-series regression (TSR) that utilized Zika search volume from Google Trends to anticipate confirmed Zika cases in Brazil and Colombia. The model, with a 1-week lag of Zika query and a 1-week lag of Zika cases as a control for autocorrelation, was found to be the most effective in predicting Zika cases. The results demonstrated the potential to forecast Zika cases a week ahead of outbreaks, offering healthcare authorities an early indicator for outbreak evaluation and precautionary measures. Using Google Trends, Ortiz-Martínez et al. [68] showed that there was a high correlation between the COVID-19 incidence in Colombia and Google searches on COVID-19 in Colombia ($R^2$ = 0.8728 and $p < 0.0001$). In addition to the above, in the last few years, Google Trends has also had a wide range of interdisciplinary applications related to the understanding and analysis of public health concerns [69–71], societal problems [72–74], emerging technologies [75–77], human behavior analysis [78–81], assistive technologies [82–85], humanitarian issues [86–89], and smart technologies [90–93]. Therefore, it may be concluded that prior works in this field have focused on using Google Trends related to mining, analysis, and investigation of multimodal components of web behavior for a wide range of applications and use cases, with a specific focus on studying and analyzing web behavior during various virus outbreaks. However, these works have multiple limitations. The following is a summary of the same and an overview of the public health needs in the context of Disease X:

- In the last few years, a significant amount of research related to the mining and analysis of web behavior, associated with different virus outbreaks such as COVID-19 [40,68,84,92], influenza [62], Lyme disease [63], malaria [64], dengue [64], chikungunya [64], syphilis [65], HIV [66], and Zika virus [67], has been conducted. Even though Disease X features in the shortlist of blueprint priority diseases of the WHO, no prior work in this field has focused on Disease X. Therefore, it is crucial to perform the mining and analysis of web behavior related to Disease X.
- The works that analyzed relevant data from Google Trends during virus outbreaks of the past have focused on web behavior originating from a very limited number of geographic regions. For example, the work by Verma et al. [64] focused on the analysis of the web behavior from two regions in India, the work by Young et al. [66] focused on web behavior analysis from the United States, and the work of Morsy et al. [67] focused on the web behavior analysis from Brazil and Columbia. Similar to the virus outbreaks of the past, which were not localized in one or two geographic regions, the outbreak of Disease X is expected to have a global impact. Therefore, the need of the hour is to mine and analyze the web behavior data related to Disease X emerging from different geographic regions.

To address these limitations, public health needs, and to contribute to the timely advancement of research and development in this field, this work presents a dataset that comprises web behavior data related to Disease X that emerged from 94 regions between February 2018 and August 2018. These 94 regions were selected for the development of this dataset as all these regions recorded a significant level of interest towards Disease X during this timeframe. This dataset was developed by collecting data from Google Trends. This paper also presents an analysis of this dataset and discusses potential applications of the same in the context of public health needs related to Disease X. The methodology that was followed for the development of this dataset is presented in Section 3.

## 3. Methodology

Google Trends [94], a tool developed by Google, allows the mining and analysis of real-time and historical information associated with Google Search queries, enabling researchers to uncover valuable insights into the interests of individuals across different domains and topics [95]. Google Trends analyzes search behavior by considering searches on Google and can thus provide unique insights associated with web-behavior. This feature is particularly valuable in health informatics, where understanding public engagement and interests in health-related topics and predicting disease outbreaks is of paramount importance [96].

The real-time data availability of Google Trends makes it superior to traditional survey methods, and it is also far less time-consuming. Additionally, as the web behavior data available via Google Trends is anonymous, it allows researchers to explore different forms of data analysis that might have been otherwise difficult due to privacy concerns of the general public [96]. Google Trends presents several significant advantages over traditional survey methods, positioning it as a potent tool for research and analysis of the multimodal characteristics of web-behavior. The foremost advantage lies in the cost-effectiveness of utilizing Google Trends. Unlike traditional surveys, which frequently entail significant expenses for participant recruitment, data collection, and analysis, Google Trends operates as a cost-free resource. This financial flexibility allows researchers to channel resources into more focused areas of investigation or allocate them toward enhancing the research process itself, promoting greater flexibility in research endeavors. Another key advantage centers around the breadth and diversity of the data captured by Google Trends. Conducting regular surveys on a global scale is a logistical challenge, often constrained by geographic and demographic limitations. However, Google Trends seamlessly aggregates web behavior data on a global scale, which can be used for in-depth study and analysis. This global perspective of Google Trends enhances the generalizability of findings and facilitates cross-cultural comparisons, making it a valuable resource for understanding the intricacies of web behavior across different geographic regions. Moreover, the near real-time nature of data availability on Google Trends is a gamechanger. Google Trends offers almost immediate access to search trends as they unfold, providing researchers with timely access to evolving interests and trends. This swift access to information enables timely analysis, decision-making, and trend detection, making it particularly advantageous in fields that require quick response, such as public health and policy formulation. In contrast, traditional surveys often grapple with time delays, influenced by the labor-intensive nature of participant recruitment and adherence to inclusion criteria. The delays inherent in survey-based research can hinder the ability to capture real-time insights, potentially impacting the accuracy and relevancy of the findings. The instant accessibility of Google Trends data addresses this limitation, empowering researchers with the agility to adapt and react promptly to emerging trends or shifts in user interests related to a topic as evidenced by relevant web-behavior.

Google Trends presents the frequency at which a specific search term is input into Google's search engine relative to the overall search volume on the site during a specific timeframe. Mathematically, if $n(q,l,t)$ represents the number of searches for the query $q$ in the location $l$ during the period $t$, the relative popularity (RP) of the query is computed as shown in Equation (1). In Equation (1), $Q(l,t)$ is a set of all the queries made from location $l$ at time $t$, $\Pi(n(q,l,t) > \tau)$ is a dummy term with a value of 1 when $n(q,l,t) > \tau$ (query is popular) and 0 otherwise. The resulting numbers are then scaled within the range of 0 to 100 based on the proportion of the topic relative to the total number of search topics. This defines the Google Trends Index (GTI), as shown in Equation (2) [96].

$$RP_{(q,l,t)} = \frac{n_{(q,l,t)}}{\Sigma_{q \epsilon Q(l,t)} n(q,l,t)} \times \Pi_{(n(q,l,t) > \tau)} \quad (1)$$

$$GTI_{(q,l,t)} = \frac{RP(q,l,t)}{max\{RP(q,l,t)_{t \epsilon 1,2,...,T}\}} \times 100 \quad (2)$$

These index values can be generated by Google Trends starting from 1 January 2004, up to 36 hours prior to the present search. Google Trends excludes search data from very limited users and highlights popular search topics while assigning 0 to terms with low search volumes [97]. The following is an overview of the features of Google Trends:

- Search Term Trends: This feature allows users to see how the popularity of a specific search term or keyword has changed over time. Google Trends provides a graphical representation to highlight these trends.
- Related Queries: Google Trends displays related queries that are frequently searched alongside the user's primary search term. This can help identify related topics or terms relevant for data analysis.
- Regional Interest: Users can view the geographical regions where a specific search term is most popular using Google Trends. Google Trends provides insights into regional differences in search interests for search terms.
- Trending Searches: This feature of Google Trends highlights the current and popular search queries or topics, providing real-time insights into what people are searching for on Google.
- Year in Search: Google Trends often releases a "Year in Search" report summarizing the top search queries from the past year. This report offers an overview of significant events and trends.
- Category Comparison: Users can compare the search interests of different categories or topics on Google using Google Trends. This can be useful for understanding the relative popularity of various topics.
- Time Period Selection: Google Trends allows users to specify the time period for which they wish to query and analyze the data. This can range from a few hours to multiple years.
- Data Visualization: Google Trends provides interactive charts and graphs to visualize search data.
- Real-Time Data: Google Trends often updates in near real-time, making it valuable for tracking ongoing events.
- Data Export: Google Trends allows different options to export data related to search interests, related queries, and related topics for a search term on Google for further analysis.

For developing this dataset, the web behavior data in terms of search interests related to Disease X (as a topic) was collected using Google Trends from February 2018 to August 2023. To perform this task, the global search trends related to Disease X (as a topic) during this timeframe (February 2018 to August 2023) were mined using Google Trends. The following represents the specific steps that were followed in this regard:

- Navigate to the "Explore" tab on Google Trends.
- Set the search query as Disease X (Topic).
- Set the geolocation to "Worldwide".
- Navigate to the timestamp dropdown menu, select "Custom Time Range", and enter the time range as "2/1/18—8/8/18".
- Set "All Categories" in the categories option.
- Select "Web Search" for the type of search.

February 2018 was selected as the start time, as the WHO added Disease X to their shortlist of blueprint priority diseases in February 2018. 8 August 2023 was the most recent date at the time of data collection. The search query was set as Disease X (Topic) to mine the search interests related to Disease X as a topic on a global scale. This selection ensured that different search queries on Google related to Disease X focused on different topics, such as *how to prepare for Disease X*, *policies to reduce the spread of Disease X*, *treatments for Disease X*, *medications for Disease X*, *available vaccines for Disease X*, *effect of Disease X on the education industry*, *effect of Disease X on the global economy*, *impact of Disease X on the healthcare industry*, *effect of Disease X on stock markets*, and *Disease X and stay-at-home guidelines*, just to name a few; these were included in the search interest values being computed. In the categories option on Google Trends, different web search categories are available. However, for the data collection, the "All Categories" option was selected to take into account all the different search categories on Google in the context of web searches about Disease X. These web search categories include *Arts and Entertainment*, *Autos and Vehicles*, *Beauty and Fitness*, *Books and Literature*, *Business and Industrial*, *Computers and Electronics*, *Finance*, *Food and Drink*, *Games*, *Health*, *Hobbies and Leisure*, *Home and Garden*, *Internet and Telecom*, *Jobs and Education*, *Law and Government*, *News*, *Online Communities*, *People and Society*, *Pets and Animals*, *Real Estate*, *Reference*, *Science*, *Shopping*, *Sports*, and *Travel*. The result provided by Google Trends is shown in Figure 1. Thereafter, by using the "Regional Interest" feature of Google Trends, the list of regions that recorded significant search interests related to Disease X during this timeframe was compiled and exported. This list of regions is shown in Table 1.

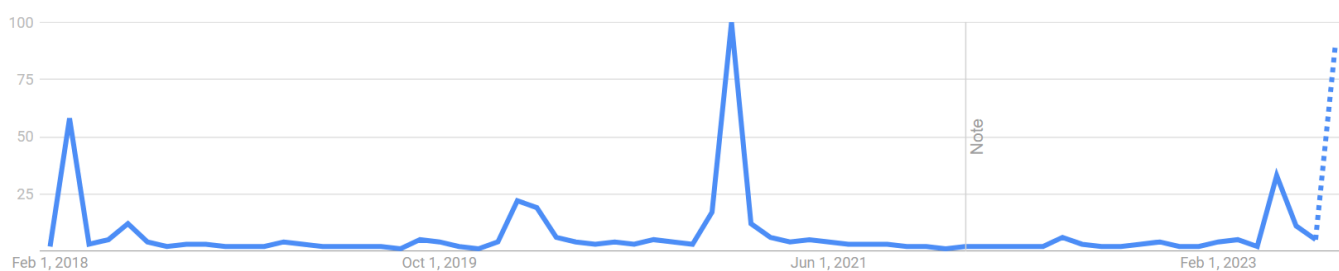

**Figure 1.** Trends in Search Interests related to Disease X (as a topic) on a Global Scale between February 2018 and August 2023.

**Table 1.** List of 94 regions that recorded significant search interests related to Disease X (as a topic) between February 2018 and August 2023.

| List of Regions |
| --- |
| Singapore, Haiti, Honduras, El Salvador, Madagascar, Panama, Bolivia, Reunion, Guatemala, Cuba, United Arab Emirates, Paraguay, Nicaragua, Hong Kong, Macao, Qatar, United Kingdom, Brunei, Ecuador, Uruguay, Oman, Bahrain, Ireland, Kuwait, Costa Rica, Argentina, India, Puerto Rico, Venezuela, France, St. Helena, Brazil, Mexico, Côte d'Ivoire, Peru, Canada, Australia, Zimbabwe, Colombia, United States, Luxembourg, Lebanon, Ghana, Algeria, New Zealand, Portugal, Malaysia, Myanmar (Burma), Ethiopia, Dominican Republic, China, Chile, Nepal, Belgium, Iraq, Taiwan, South Africa, Tunisia, Sri Lanka, Thailand, Switzerland, Spain, Bangladesh, Saudi Arabia, Kenya, South Korea, Germany, Norway, Pakistan, Indonesia, Hungary, Morocco, Austria, Israel, Nigeria, Bulgaria, Philippines, Netherlands, Denmark, Greece, Italy, Jordan, Egypt, Sweden, Finland, Czechia, Romania, Poland, Iran, Türkiye, Russia, Vietnam, Ukraine, Japan |

Thereafter, by utilizing Google Trends as the data source, search interests related to Disease X (as a topic) for all these 94 regions between February 2018 and August 2023, were collected and exported as .CSV files. As far as other geographic regions are concerned, for instance, Yemen, Zimbabwe, Tajikistan, Namibia, Fiji, etc., there was not a significant number of Google searches related to Disease X on a monthly basis between February 2018 and August 2023. As a result, Google Trends did not provide any value for search interests related to Disease X from all such regions. Therefore, such regions were not included in the dataset development. To consolidate the 94 .CSV files into one workbook on Microsoft Excel, the Power Query interface on Excel was employed. The Power Query tool used each individual file as a data source and imported each file's data into the Excel Workbook. Each region's search interest for "Disease X" is present as a different sheet in this file, which was uploaded to IEEE Dataport [98] as a dataset. The flowchart in Figure 2 shows the step-by-step process that was followed for the development of this dataset. This dataset is described in Section 3.

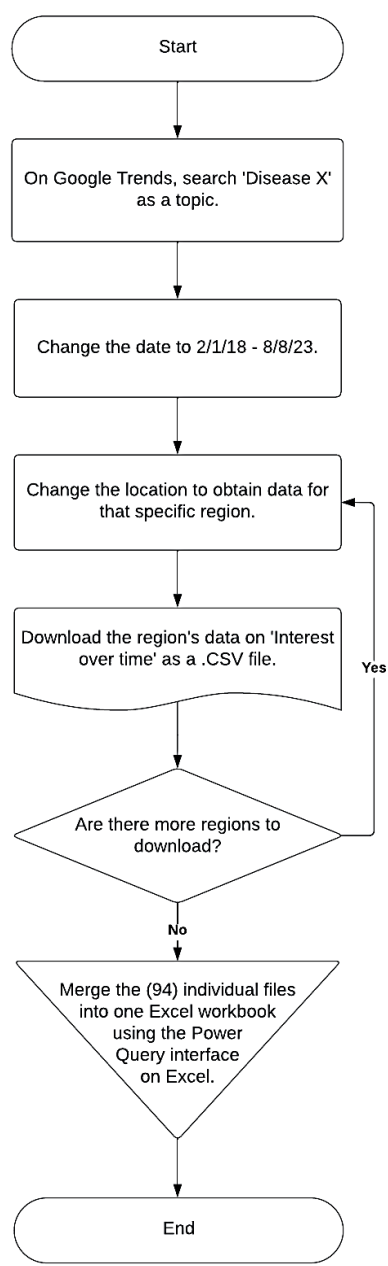

**Figure 2.** A flowchart to represent the step-by-step process of the development of this dataset.

## 4. Data Description

This section describes the dataset, which is available at https://dx.doi.org/10.21227/ht7f-rx42. This dataset contains one Microsoft Excel workbook that comprises 94 different sheets, where each sheet presents the search interests related to Disease X (as a topic) for a different region between February 2018 and August 2023. The search interest data for all the regions stated in Table 1 is available in this dataset. For each region, this dataset presents the search interests related to Disease X (as a topic) for each month in this timeframe, i.e., from February 2018 to August 2023. Table 2 presents the description of the different attributes present in this dataset. It is worth mentioning here that these attributes are present across different sheets in the dataset, where each sheet contains the name of the region. For instance, in the sheet named—"Singapore", the two attributes present are "Month" and "Disease X: (Singapore)". Similarly, in the sheet named "Honduras", the two attributes present are "Month" and "Disease X: (Honduras)". There are 94 such sheets in this dataset, which are named as per these regions. To avoid presenting a table with 95 rows, all the attributes from all the sheets are not listed in Table 2.

**Table 2.** Description of different attributes present in the dataset.

| Attribute Name | Description | Range | Datatype |
|---|---|---|---|
| Month | Represents each month between February 2018 and August 2023 | February 2018–August 2023 | Date |
| Disease X: (Singapore) | Search Interest Data about Disease X from Singapore | 0–100 | Numerical |
| Disease X: (Honduras) | Search Interest Data about Disease X from Honduras | 0–100 | Numerical |
| Disease X: (Haiti) | Search Interest Data about Disease X from Haiti | 0–100 | Numerical |
| Disease X: (Nicaragua) | Search Interest Data about Disease X from Nicaragua | 0–100 | Numerical |
| Disease X: (Guatemala) | Search Interest Data about Disease X from Guatemala | 0–100 | Numerical |
| Disease X: (El Salvador) | Search Interest Data about Disease X from El Salvador | 0–100 | Numerical |
| Disease X: (Brunei) | Search Interest Data about Disease X from Brunei | 0–100 | Numerical |
| Disease X: (Panama) | Search Interest Data about Disease X from Panama | 0–100 | Numerical |
| . | . | . | . |
| . | . | . | . |
| . | . | . | . |
| Disease X: (Japan) | Search Interest Data about Disease X from Japan | 0–100 | Numerical |
| Disease X: (Ukraine) | Search Interest Data about Disease X from Ukraine | 0–100 | Numerical |
| Disease X: (Vietnam) | Search Interest Data about Disease X from Vietnam | 0–100 | Numerical |

In the remainder of this section, the compliance of this dataset with the FAIR principles of Scientific Data Management [99] is explained. The FAIR principles include four key aspects of scientific data management, namely Findability, Accessibility, Interoperability, and Reusability. The components of the FAIR Principles exhibit interrelationships while maintaining autonomy and distinctiveness. The aforementioned principles delineate specific factors to be taken into account in modern data publication settings, particularly in relation to facilitating both human and computerized methods of depositing, exploring, accessing, collaborating, and reusing data. In the last few years, numerous works have emerged, primarily focused on specific domains, advocating for enhancements in the handling of data and data archival practices. However, FAIR stands apart by presenting succinct, domain-agnostic, overarching principles that can be universally applied to diverse research results. The principles delineate the essential attributes that modern data resources, tools, vocabularies, and infrastructures should possess in order to facilitate the process of identification and enable the reuse of such resources by others. The principles may be followed in various configurations and progressively, as data providers' publication settings progress towards higher levels of 'FAIRness'. Furthermore, the flexibility of the principles, together with their clear differentiation between data and metadata, provides explicit support for a diverse array of scenarios. The FAIR Guiding Principles, which are of a higher level, come before the selection of implementation options and do not endorse any particular technology, standard, or implementation approach. It is important to note that these principles do not constitute a standard or a specification in and of themselves. These guidelines serve as a reference for data publishers and stewards, aiding them in assessing the effectiveness of the resulting decisions in ensuring that their research artifacts are defined by the principles of Findability, Accessibility, Interoperability, and Reusability. These overarching principles facilitate a diverse array of integrated and investigative behaviors, grounded in a vast selection of technological options and applications. The FAIR principles comprise the potential to have beneficial effects on a wide range of stakeholders. These include researchers who seek to distribute, receive recognition for, and utilize one another's data and solutions. Essentially, the FAIR principles endeavor to cultivate a more cooperative and transparent research landscape, facilitating the exchange of knowledge and bolstering the lasting influence of scientific investigations related to database development and database management [99]. Several prior works in the field of dataset development have discussed how developed datasets, such as the human metabolome database for 2022 [100], WikiPathways dataset [101], datasets of Tweets about COVID-19 [102,103], a dataset of Tweets about MPox [104], computational 2D materials database (C2DB) [105], the open reaction database [106], RCSB Protein Data Bank [107], and PHI-base (pathogen–host interactions database) [108], just to name a few, comply with the FAIR principles of scientific data management. This dataset, available at https://dx.doi.org/10.21227/ht7f-rx42, is findable, as it has a unique and permanent DOI assigned by IEEE Dataport. This DOI can be used by researchers from any discipline to find this dataset online. This dataset satisfies the accessibility property, as it can be accessed by any user on the internet using any device via the DOI, as long as the user's device is connected to the internet and is operating in a desired manner. The dataset is interoperable, as the data in this dataset is available in a standard format (.xlsx file) that can be downloaded, read, and analyzed across different computer systems, frameworks, and applications. Finally, this dataset satisfies the reusability property as the data can be re-used any number of times for the study and investigation of different research questions that focus on the analysis of search interests related to Disease X.

## 5. Data Analysis and Potential Applications

This section presents the results obtained from a brief analysis of this dataset. It concludes by highlighting a few potential applications of this dataset. The data present in this dataset can be analyzed to obtain the trends in search interests during this timeframe for each of these 94 regions. For instance, the analysis of this data for the United States is presented in Figure 3. In this Figure, the X-axis represents the months, and the Y-axis represents the search interest related to Disease X on a scale of 0 to 100.

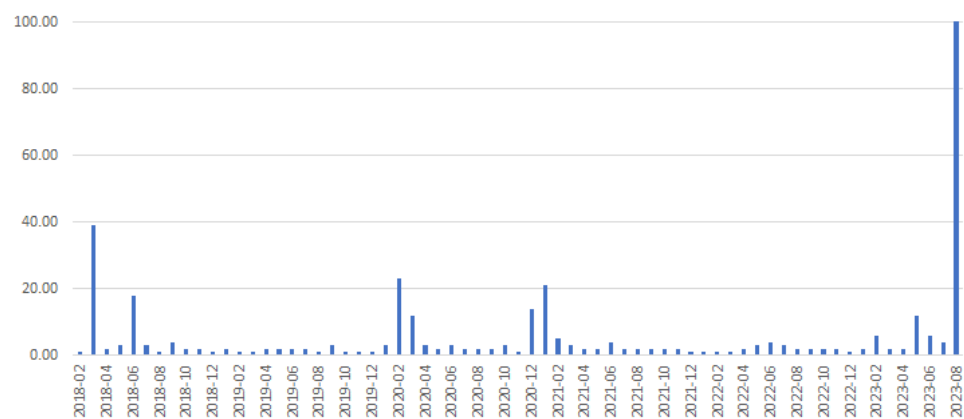

**Figure 3.** Trends in search interests related to Disease X (as a topic) for the United States between February 2018 and August 2023.

This analysis of the data for the United States shows that the search interest related to Disease X was the highest in August 2023. Similar trends and insights associated with search interests for Disease X emerging from different geographic regions can be obtained from analysis of the search interest data for that region as available in this dataset. The data present in this dataset also unravel the evolving paradigms of global search patterns about Disease X. For instance, Figure 4 shows a world map-based representation of different regions that recorded a significant number of Google searches about Disease X on 8 August 2023. As can be seen from this Figure, these regions were Fiji, Oman, Ethiopia, the United Kingdom, Uganda, Uruguay, El Salvador, Puerto Rico, Nepal, Canada, Ecuador, Venezuela, Bangladesh, Morocco, Bolivia, Singapore, Bulgaria, South Africa, Ireland, the United States, the United Arab Emirates, Australia, the Netherlands, Hong Kong, Israel, Nigeria, Austria, Belgium, India, Pakistan, Portugal, Colombia, Egypt, Argentina, France, Poland, Thailand, Malaysia, Spain, Germany, and Russia. Figure 5 shows a world map-based representation of different regions that recorded a significant number of Google searches about Disease X on 8 July 2023. As can be seen from Figure 5, the regions that recorded a significant number of Google searches about Disease X on 8 July 2023 were Maldives, Cuba, Kuwait, Morocco, the Dominican Republic, Bulgaria, Costa Rica, Uruguay, New Zealand, Israel, Pakistan, Peru, the United Arab Emirates, Greece, Philippines, Argentina, Thailand, Spain, Türkiye, Brazil, and the United States. These two figures illustrate the fact that the global landscape, in terms of interest in Disease X, significantly changed over a time period of just one month, between 8 July 2023 and 8 August 2023.

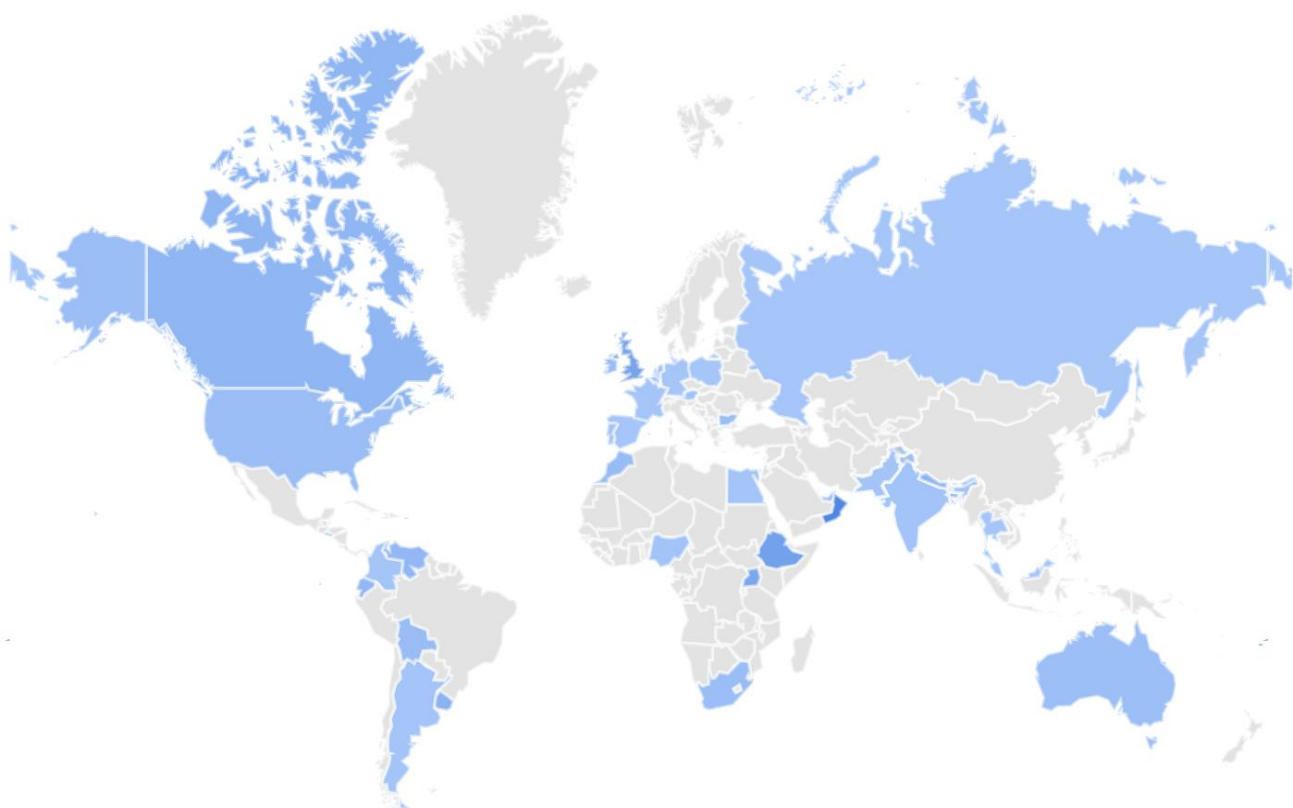

**Figure 4.** A world map-based analysis of the significant number of Google searches related to Disease X from different regions of the world on 8 August 2023.

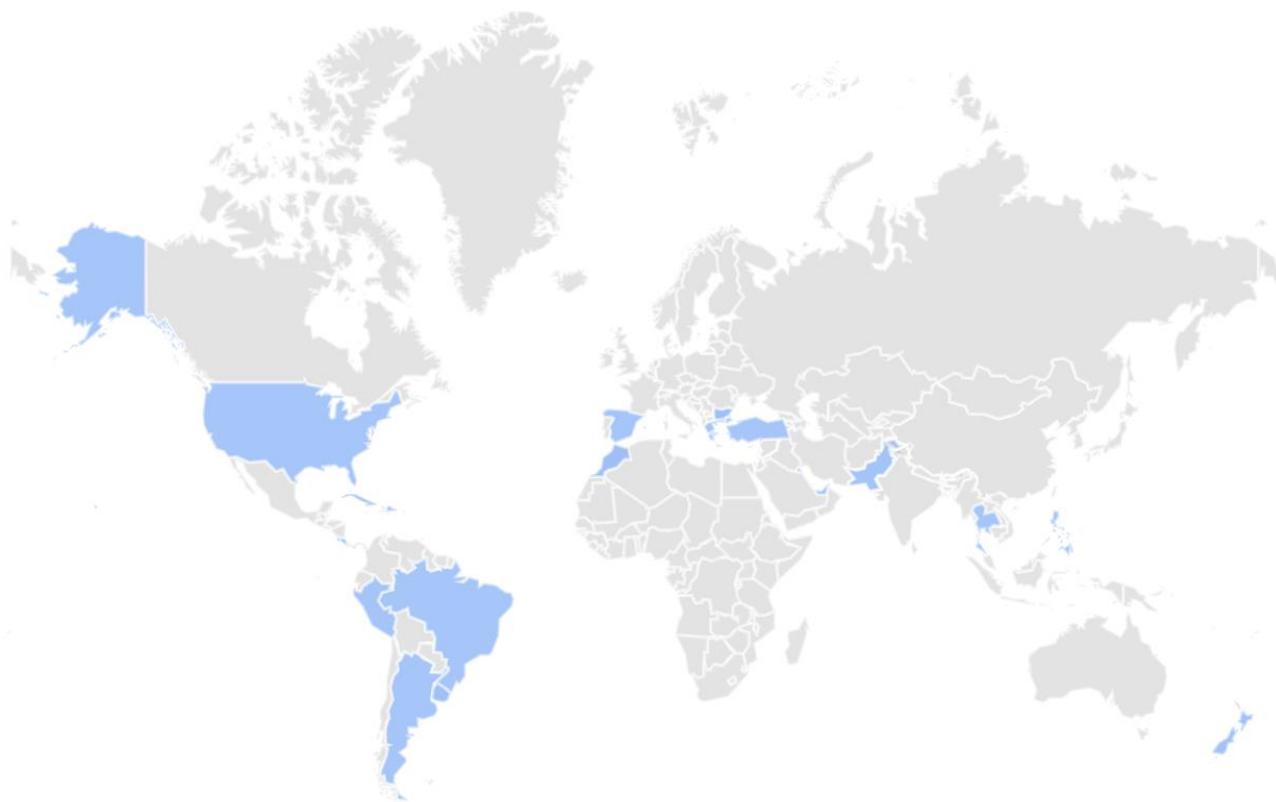

**Figure 5.** A world-map-based analysis of the significant number of Google searches related to Disease X from different regions of the world on 8 July 2023.

Figures 4 and 5 also indicate that within a short period of time, people from several regions of the world have started proactively searching for Disease X on Google. In a similar manner, the evolution of global interest related to Disease X over different time periods can be mapped, analyzed, and interpreted using this dataset.

During the development of this dataset, it was observed that online searches on Google related to Disease X during this timeframe (February 2018 to August 2023) had several related queries. The 'rising' keywords associated with these related queries were collected using the "Related Queries" feature of Google Trends, as described in Section 2. Figure 6 shows a word-cloud-based representation of these queries related to Disease X during this timeframe. In this context, it is worth mentioning that the mining of the data from Google Trends for the development of this dataset was performed on 8 August 2023. Google Trends provided the search interest for August 2023 for each of the 94 regions by taking into account the relevant Google Searches recorded from 1 August 2023 to 8 August 2023. So, if the data collection is performed once again at the end of August 2023 or at a later date using Google Trends, it is possible that the search interest for August 2023 for some of these regions might change, as Google Trends would then report the search interest value for August 2023 by taking into account all relevant Google Searches recorded from 1 August 2023 to 31 August 2023.

Thereafter, a comprehensive analysis of the search interests associated with Disease X from all 94 regions between February 2018 and August 2018 was performed to explore and investigate the trends of the same. These are presented in Figures 7–15. Multiple graphical representations were prepared, primarily to ensure the readability of the compared trends for investigation of the underlying search interests. Each of these graphs presents the trends in search interests about Disease X for about 10 distinct countries, enabling the exploration and investigation of the trends in search interests. For instance, from Figure 7 it can be inferred that the number of Google Searches about Disease X in Brunei in August 2023 was much higher than the number of Google Searches about Disease X from Singapore, Honduras, Haiti, Nicaragua, Guatemala, El Salvador, Panama, Cuba, and the United Arab Emirates.

**Figure 6.** A word-cloud-based representation of 'rising' queries related to Disease X from February 2018 to August 2023.

**Figure 7.** A graphical analysis of search interests (monthly) related to Disease X in Singapore, Honduras, Haiti, Nicaragua, Guatemala, El Salvador, Brunei, Panama, Cuba, and the United Arab Emirates between February 2018 and August 2023.

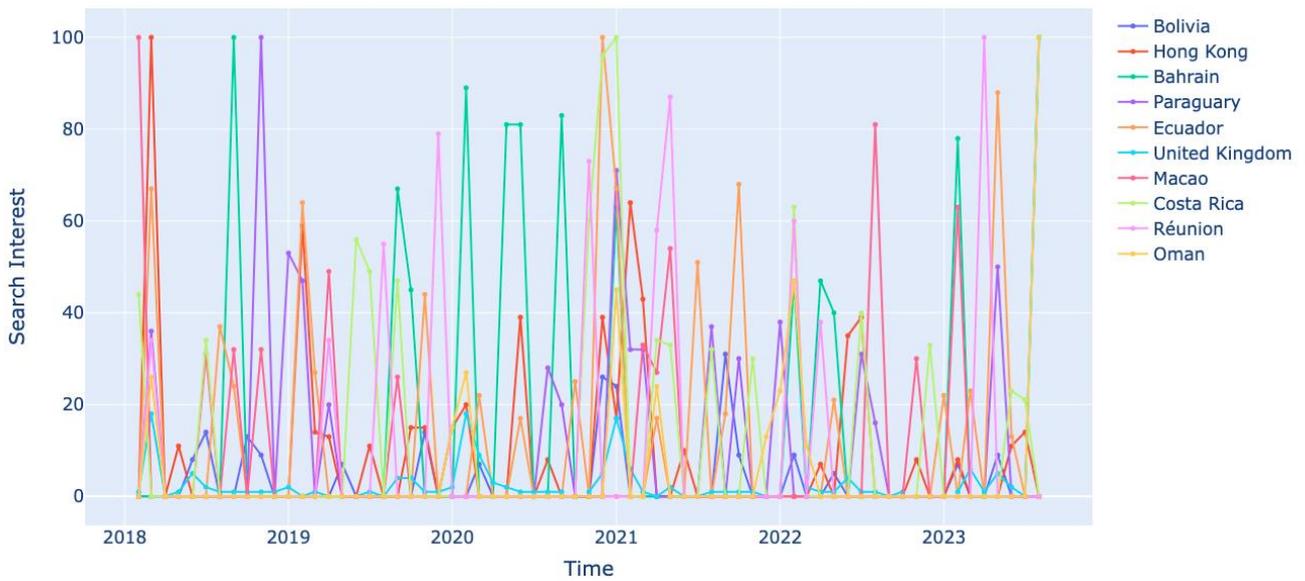

**Figure 8.** A graphical analysis of search interests (monthly) related to Disease X in Bolivia, Hong Kong, Bahrain, Paraguay, Ecuador, the United Kingdom, Macao, Costa Rica, Reunion, and Oman between February 2018 and August 2023.

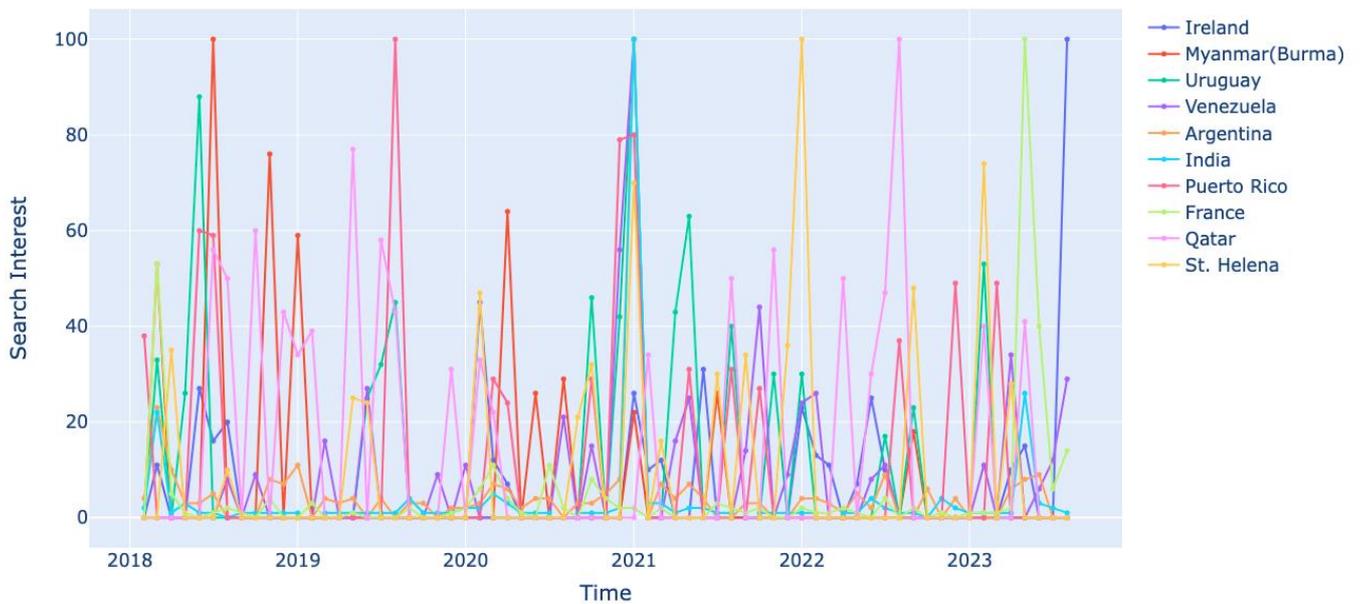

**Figure 9.** A graphical analysis of search interests (monthly) related to Disease X in Ireland, Myanmar (Burma), Uruguay, Venezuela, Argentina, India, Puerto Rico, France, Qatar, and St. Helena between February 2018 and August 2023.

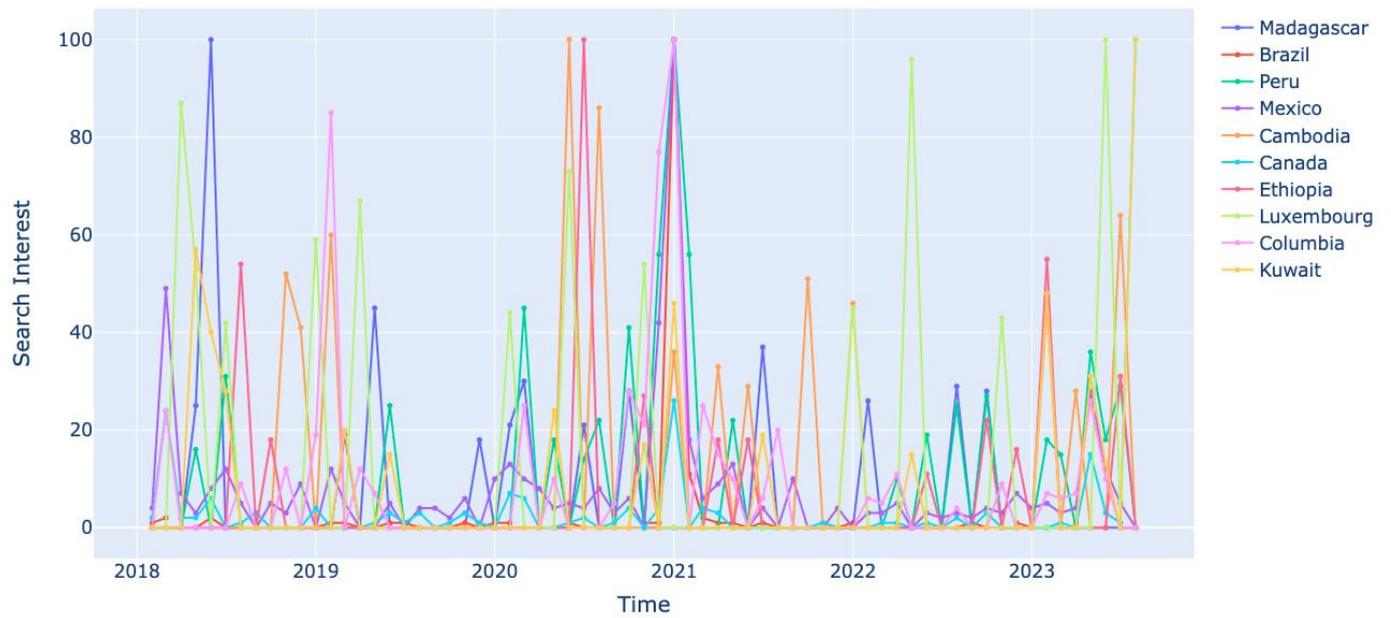

**Figure 10.** A graphical analysis of search interests (monthly) related to Disease X in Madagascar, Brazil, Peru, Mexico, Cambodia, Canada, Ethiopia, Luxembourg, Colombia, and Kuwait between February 2018 and August 2023.

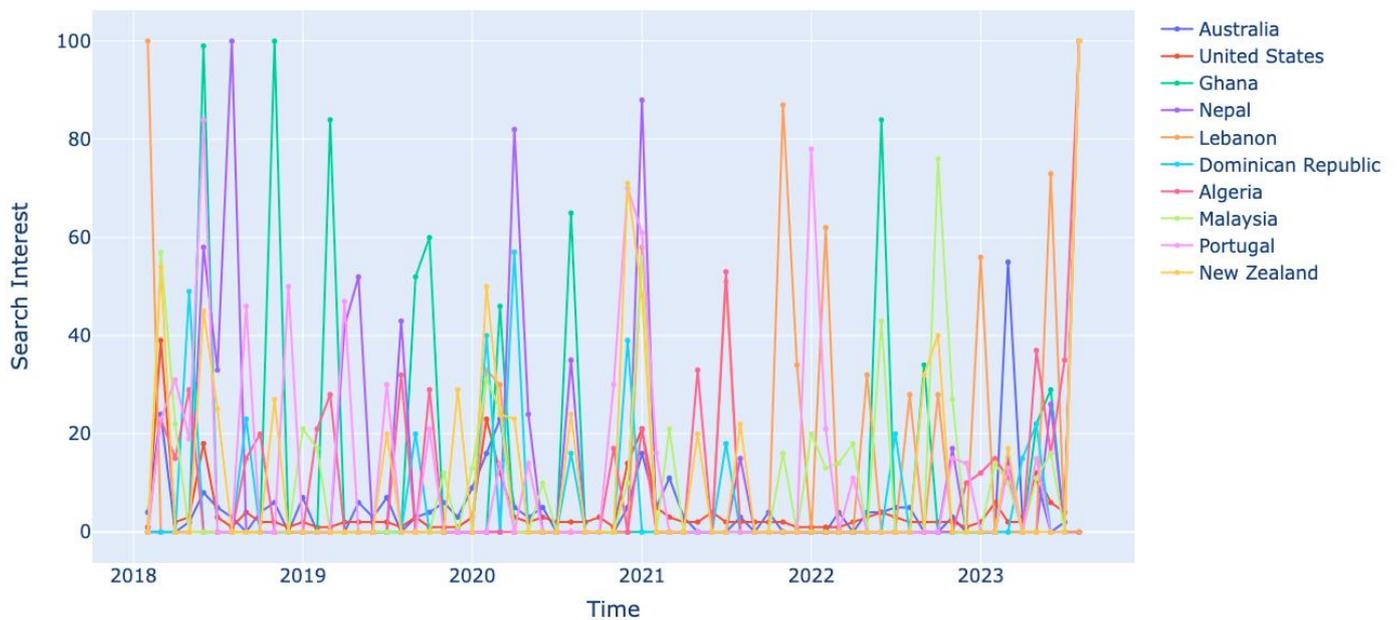

**Figure 11.** A graphical analysis of search interests (monthly) related to Disease X in Australia, the United States, Ghana, Nepal, Lebanon, the Dominican Republic, Algeria, Malaysia, Portugal, and New Zealand, between February 2018 and August 2023.

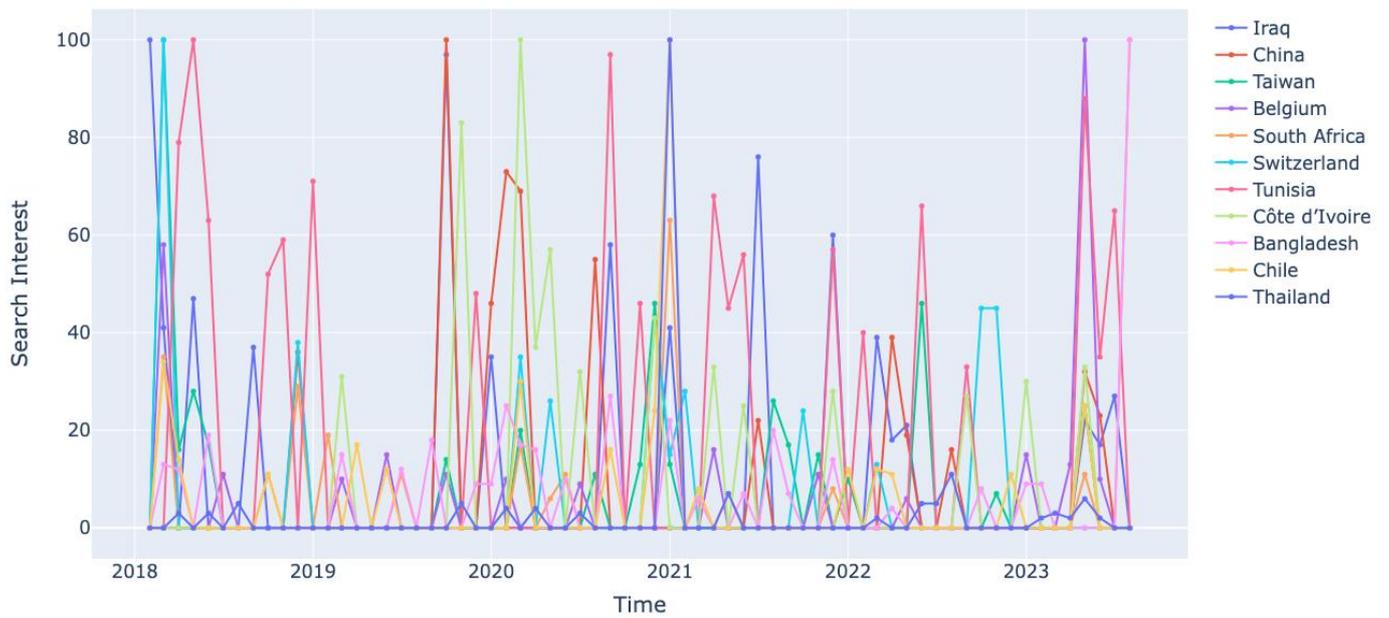

**Figure 12.** A graphical analysis of search interests (monthly) related to Disease X in Iraq, China, Taiwan, Belgium, South Africa, Switzerland, Tunisia, Côte d'Ivoire, Bangladesh, Chile, and Thailand between February 2018 and August 2023.

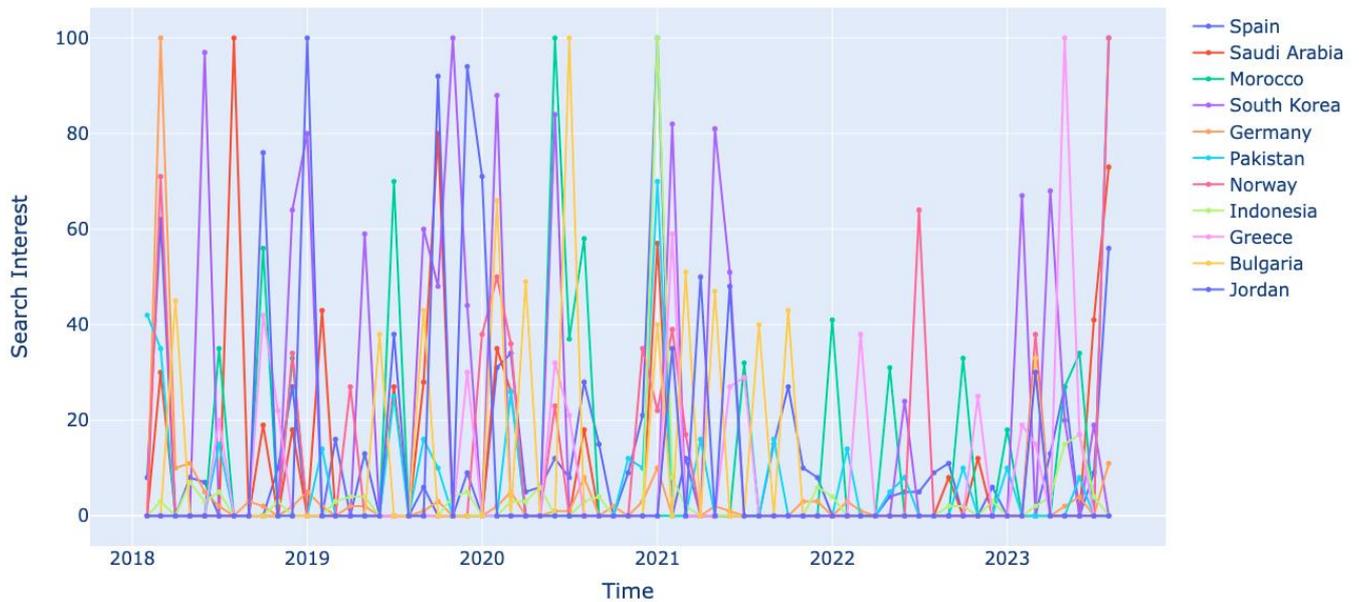

**Figure 13.** A graphical analysis of search interests (monthly) related to Disease X in Spain, Saudi Arabia, Morocco, South Korea, Germany, Pakistan, Norway, Indonesia, Greece, Bulgaria, and Jordan between February 2018 and August 2023.

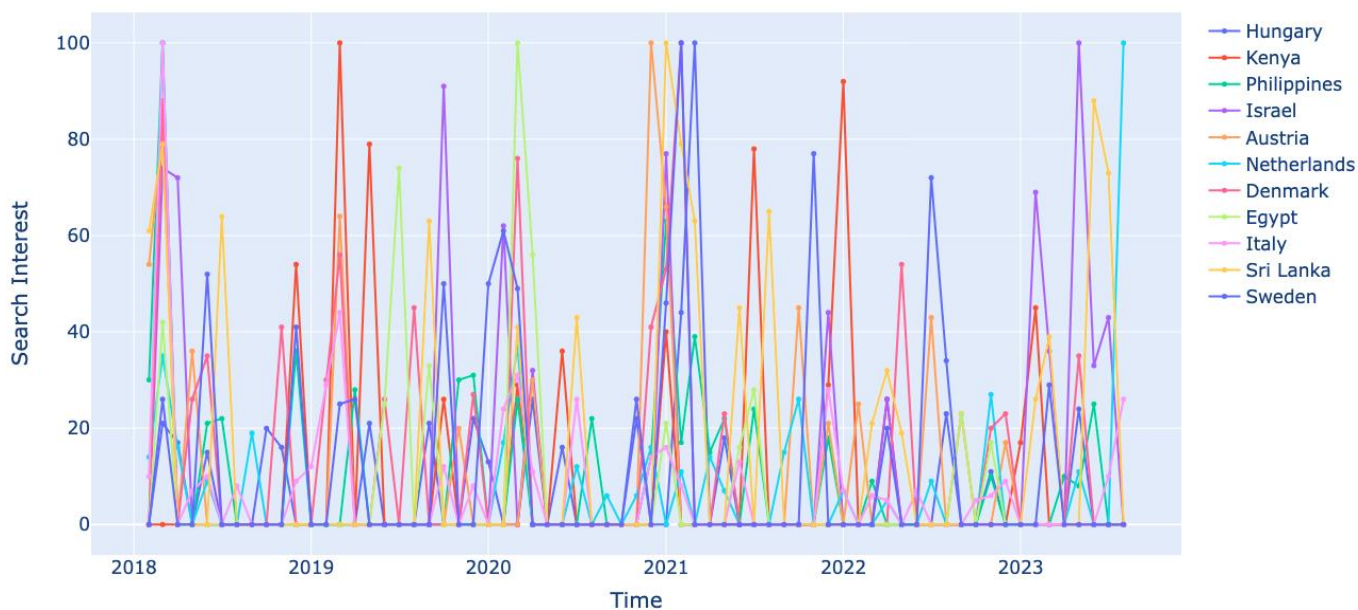

**Figure 14.** A graphical analysis of search interests (monthly) related to Disease X in Hungary, Kenya, Philippines, Israel, Austria, Netherlands, Denmark, Egypt, Italy, Sri Lanka, and Sweden between February 2018 and August 2023.

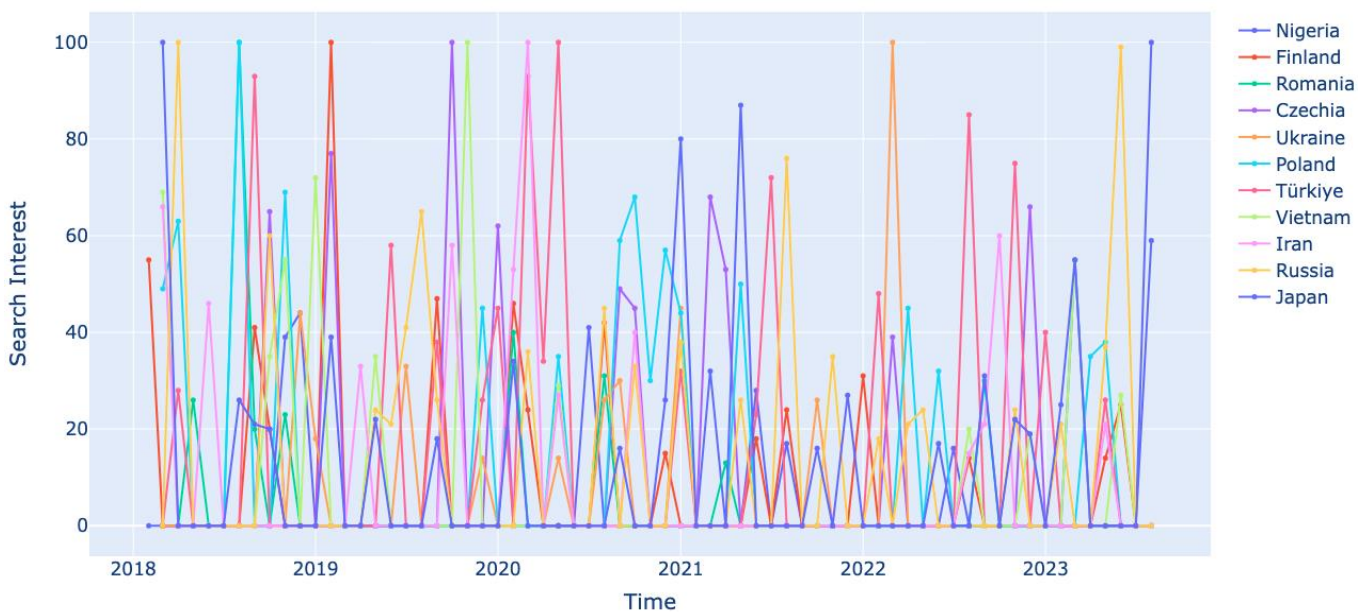

**Figure 15.** A graphical analysis of search interests (monthly) related to Disease X in Nigeria, Finland, Romania, Czechia, Ukraine, Poland, Türkiye, Vietnam, Iran, Russia, and Japan between February 2018 and August 2023.

Thereafter, a country-specific analysis of the highest search interests related to Disease X was performed. For readability, the results are shown as two different graphs in Figures 16 and 17, respectively. Multiple novel insights related to the volume of Google Searchers originating from different parts of the world can be inferred from Figures 16 and 17. For instance, a high number of Google Searches about Disease X (indicated by the search interest value being 100) was observed in several countries for August 2023. These specific regions were Bolivia, the United Kingdom, Oman, Ireland, Canada, Kuwait, Australia, the United States, the Dominican Republic, Algeria, Malaysia, Portugal, New Zealand, South Africa, Bangladesh, Pakistan, Norway, Netherlands, and Nigeria. This is a significant increase as compared to recent historical data as far as a high number of Google Searches (indicated by the search interest value being 100) related to Disease X are concerned. For instance, in June 2023, only one country, Luxembourg recorded a high number of Google Searches about Disease X (indicated by the search interest value being 100). Similarly, in April 2023, only one country, Réunion, recorded a high number of Google Searches about Disease X (indicated by the search interest value being 100). These findings indicate that there has been a considerable increase in Google Searches about Disease X from different regions of the world.

This dataset is expected to contribute towards the investigation of a wide range of research questions in different disciplines, such as Healthcare, Epidemiology, Big Data, Data Science, and Data Analysis, with a specific focus on Disease X. Given the potential correlation between the volume of internet searches and the information needs of users [109,110], this dataset could be utilized by public health organizations at both local and global levels to obtain insights into the specific needs of the general public pertaining to Disease X in various geographic regions. The use of Google Trends as a supplementary tool to conventional surveillance systems for monitoring epidemics has proven to be effective, as demonstrated by multiple prior works in this field [111,112]. So, if an epidemic due to Disease X were to start, this dataset is expected to serve as a framework for public health organizations for the development of a surveillance system for Disease X. Previous studies in this field have utilized data acquired from Google Trends to identify instances of "panic-induced searching", where media coverage of a specific outbreak amplified web-search activity [113]. For example, during the avian influenza outbreak that occurred between 2005 and 2006 [114], starting in China and spreading to Turkey, there were notable spikes in the US search volume index for the term "bird flu", despite no confirmed cases of avian flu in the US. So, the data available in the dataset could be analyzed to infer whether the spikes in search interests related to Disease X emerging from different regions were caused as a result of media coverage about Disease X or whether people in different regions of the world (that recorded spikes in search interests about Disease X) were genuinely concerned about Disease X.

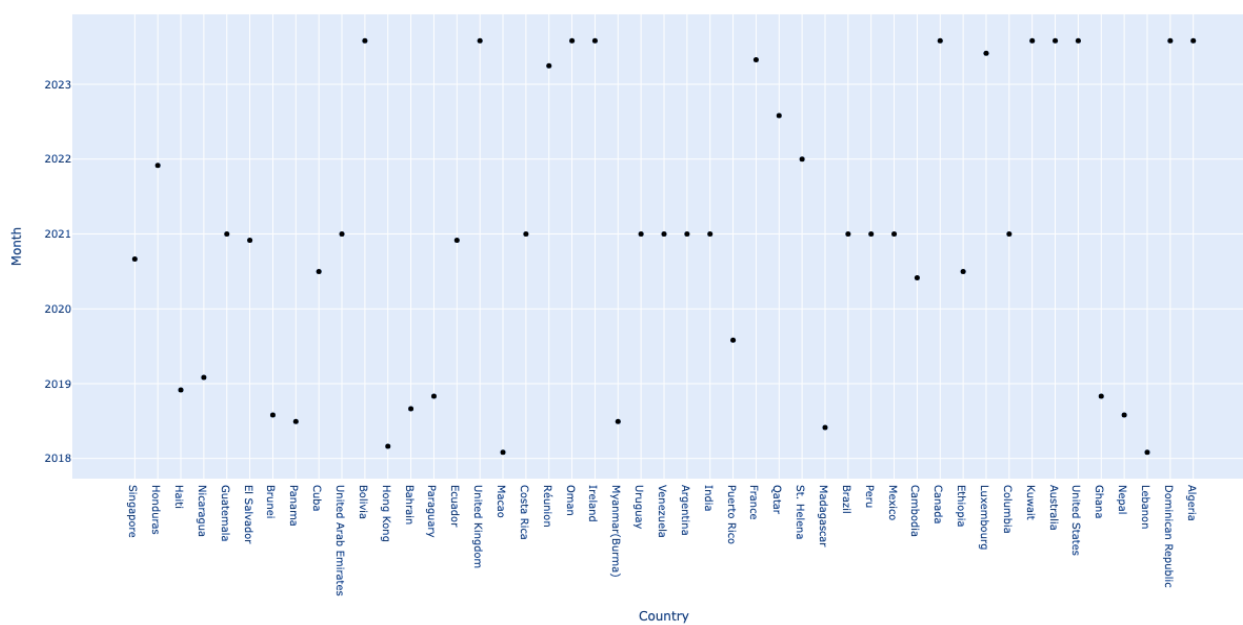

**Figure 16.** A graphical representation of the specific months between February 2018 and August 2023 when the highest search interests related to Disease X were recorded in Singapore, Honduras, Haiti, Nicaragua, Guatemala, El Salvador, Brunei, Panama, Cuba, the United Arab Emirates, Bolivia, Hong Kong, Bahrain, Paraguay, Ecuador, the United Kingdom, Macao, Costa Rica, Reunion, Oman, Ireland, Myanmar (Burma), Uruguay, Venezuela, Argentina, India, Puerto Rico, France, Qatar, St. Helena, Madagascar, Brazil, Peru, Mexico, Cambodia, Canada, Ethiopia, Luxembourg, Colombia, Kuwait, Australia, the United States, Ghana, Nepal, Lebanon, the Dominican Republic, and Algeria.

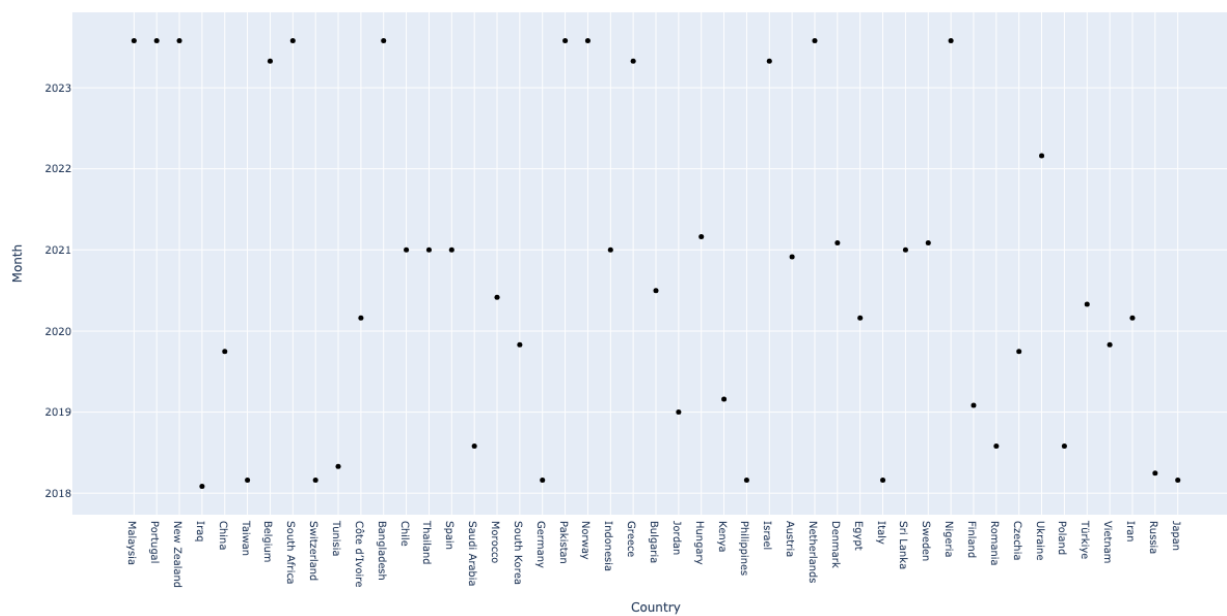

**Figure 17.** A graphical representation of the specific months between February 2018 and August 2023 when the highest search interests related to Disease X were recorded in Malaysia, Portugal, New Zealand, Iraq, China, Taiwan, Belgium, South Africa, Switzerland, Tunisia, Côte d'Ivoire, Bangladesh, Chile, Thailand, Spain, Saudi Arabia, Morocco, South Korea, Germany, Pakistan, Norway, Indonesia, Greece, Bulgaria, Jordan, Hungary, Kenya, Philippines, Israel, Austria, Netherlands, Denmark, Egypt, Italy, Sri Lanka, Sweden, Nigeria, Finland, Romania, Czechia, Ukraine, Poland, Türkiye, Vietnam, Iran, Russia, and Japan.

The internet serves as a platform for public health organizations to efficiently and affordably distribute healthcare information. However, it is crucial that reliable news is disseminated to the general public [115–117]. In the last few weeks, many public health organizations have disseminated information pertaining to Disease X, with the aim of reaching a wide audience [118–121]. The dataset presented in this work could be helpful for public health organizations to build a framework to understand the translation gap of such information about Disease X, i.e., the gap between what people need to know and what most people believe they know. Furthermore, this dataset, along with relevant social media data about Disease X, may be utilized by both global and local health organizations to identify the information requirements related to Disease X, barriers to preventing infection stemming from social and behavioral factors, and instances of misinformation about Disease X in different regions of the world. Finally, the investigation of the temporal pattern of query rates about Disease X from this dataset, in tandem with their geographical dispersion and primary search themes, may provide a measurable and meaningful indicator of the level of public interest and information requirements regarding Disease X.

The work presented in this paper has a few limitations. First, the data collected by Google Trends is limited to the search patterns emerging from only a subset of the global population, specifically those who have the ability to access the internet and who choose to use Google as their preferred search engine, as opposed to other search engines. Second, a significant constraint of Google Trends is the lack of thorough information about the methodology used by Google for generating search interest data and the algorithms utilized for its analysis. Third, the data from Google Trends available in this dataset represents relative search volumes and not absolute values of the number of Google Searches, as Google Trends only provides the relative search volume data. Finally, there is a lack of documentation on the past developments related to the design and functionalities of Google Trends. This absence of documentation may result in fluctuations in search results and therefore impact the outcomes of research studies, depending on when they were conducted.

## 6. Conclusions

The World Health Organization (WHO) added "Disease X" to their shortlist of blueprint priority diseases to represent a hypothetical, unknown pathogen that could cause a future epidemic. Since then, several works in this field have analyzed virus outbreaks of the past to propose approaches, methodologies, principles, and guidelines for better awareness, preparedness, and response towards Disease X. Many of these works that focused on analyzing virus outbreaks of the past, such as COVID-19, Influenza, Lyme disease, and Zika virus, utilized Google Trends to mine and analyze multimodal components of web behavior. However, two primary limitations exist in these works. First, these works did not specifically focus on Disease X. Second, many of these works focused on the analysis of Google Trends data originating from a very limited number of geographic regions. To address these limitations and to contribute towards the timely advancement of research in this field, this work presents a dataset of search interests related to Disease X (as a topic) originating from 94 regions of the world between February 2018 and August 2023. These 94 regions were selected for the development of this dataset as all these regions recorded a significant level of search interest towards Disease X during this timeframe. The dataset is available at https://dx.doi.org/10.21227/ht7f-rx42. In this dataset, for every region, the search interest related to Disease X is available for each month during this timeframe. The dataset complies with the FAIR principles of scientific data management. This paper also presents a brief analysis of this dataset to uphold its relevance and usefulness for the investigation of different research questions in the interrelated fields of Big Data, Data Mining, Healthcare, Epidemiology, Information Retrieval, and Data Analysis with a specific focus on Disease X. As per the best knowledge of the authors, no similar work in this field has been done so far. Future work in this area would involve analyzing the specific trends of search interests related to Disease X across different geographic regions to determine and investigate specific similarities or dissimilarities of those trends.

**Author Contributions:** Conceptualization, N.T.; methodology, N.T., K.A.P., S.C., and Y.N.D.; software, N.T., S.C., and K.A.P.; validation, N.T. and S.C.; formal analysis, N.T. and S.C.; investigation, N.T.; resources, N.T.; data curation, N.T. and K.A.P.; writing—original draft preparation, N.T., I.H., Y.N.D., K.A.P. and S.C.; writing—review and editing, N.T.; visualization, S.C. and N.T.; supervision, N.T.; project administration, N.T.; funding acquisition, not applicable. All authors have read and agreed to the published version of the manuscript.

**Funding:** This research received no external funding.

**Institutional Review Board Statement:** Not applicable.

**Informed Consent Statement:** Not applicable.

**Data Availability Statement:** This work resulted in the creation of a dataset that is available at https://dx.doi.org/10.21227/ht7f-rx42, as per the CC BY 4.0 License.

**Conflicts of Interest:** The authors declare no conflicts of interest.

## References

1. Fauci, A.S.; Lane, H.C.; Redfield, R.R. Covid-19—Navigating the Uncharted. *N. Engl. J. Med.* **2020**, *382*, 1268–1269. https://doi.org/10.1056/nejme2002387.
2. Prentice, M.B.; Rahalison, L. Plague. *Lancet* **2007**, *369*, 1196–1207. https://doi.org/10.1016/s0140-6736(07)60566-2.
3. Aassve, A.; Alfani, G.; Gandolfi, F.; Le Moglie, M. Epidemics and Trust: The Case of the Spanish Flu. *Health Econ.* **2021**, *30*, 840–857. https://doi.org/10.1002/hec.4218.
4. Joint United Nations Programme on HIV/AIDS. *World Health Organization 2008 Report on the Global AIDS Epidemic*; World Health Organization: Genève, Switzerland, 2008; ISBN 9789291737116.
5. Jacob, S.T.; Crozier, I.; Fischer, W.A., II; Hewlett, A.; Kraft, C.S.; de la Vega, M.-A.; Soka, M.J.; Wahl, V.; Griffiths, A.; Bollinger, L.; et al. Ebola Virus Disease. *Nat. Rev. Dis. Primers* **2020**, *6*, 13. https://doi.org/10.1038/s41572-020-0147-3.
6. Thakur, N.; Duggal, Y.N.; Liu, Z. Analyzing Public Reactions, Perceptions, and Attitudes during the MPox Outbreak: Findings from Topic Modeling of Tweets. *Computers* **2023**, *12*, 191. https://doi.org/10.3390/computers12100191.
7. Sampath, S.; Khedr, A.; Qamar, S.; Tekin, A.; Singh, R.; Green, R.; Kashyap, R. Pandemics throughout the History. *Cureus* **2021**, *13*, e18136. https://doi.org/10.7759/cureus.18136.

8. Chatterjee, P.; Nair, P.; Chersich, M.; Terefe, Y.; Chauhan, A.; Quesada, F.; Simpson, G. One Health, "Disease X" & the Challenge of "Unknown" Unknowns. *Indian J. Med. Res.* **2021**, *153*, 264. https://doi.org/10.4103/ijmr.ijmr_601_21.
9. Prioritizing Diseases for Research and Development in Emergency Contexts. Available online: https://www.who.int/activities/prioritizing-diseases-for-research-and-development-in-emergency-contexts (accessed on 17 August 2023).
10. Barnes, T. World Health Organisation Fears New "Disease X" Could Cause a Global Pandemic. Available online: https://www.independent.co.uk/news/science/disease-x-what-is-infection-virus-world-health-organisation-warning-ebola-zika-sars-a8250766.html (accessed on 17 August 2023).
11. Scutti, S. World Health Organization Gets Ready for 'Disease X'. Available online: https://www.cnn.com/2018/03/12/health/disease-x-blueprint-who/index.html (accessed on 17 August 2023).
12. Adalja, A.A.; Watson, M.; Toner, E.S.; Cicero, A.; Inglesby, T.V. Characteristics of Microbes Most Likely to Cause Pandemics and Global Catastrophes. In *Current Topics in Microbiology and Immunology*; Springer International Publishing: Cham, Switzerland, 2019; pp. 1–20; ISBN 9783030363109.
13. Kreuder Johnson, C.; Hitchens, P.L.; Smiley Evans, T.; Goldstein, T.; Thomas, K.; Clements, A.; Joly, D.O.; Wolfe, N.D.; Daszak, P.; Karesh, W.B.; et al. Spillover and Pandemic Properties of Zoonotic Viruses with High Host Plasticity. *Sci. Rep.* **2015**, *5*, 14830. https://doi.org/10.1038/srep14830.
14. Jones, K.E.; Patel, N.G.; Levy, M.A.; Storeygard, A.; Balk, D.; Gittleman, J.L.; Daszak, P. Global Trends in Emerging Infectious Diseases. *Nature* **2008**, *451*, 990–993. https://doi.org/10.1038/nature06536.
15. Carlson, C.J.; Albery, G.F.; Merow, C.; Trisos, C.H.; Zipfel, C.M.; Eskew, E.A.; Olival, K.J.; Ross, N.; Bansal, S. Climate Change Increases Cross-Species Viral Transmission Risk. *Nature* **2022**, *607*, 555–562. https://doi.org/10.1038/s41586-022-04788-w.
16. Peiris, J.S.M.; Tu, W.-W.; Yen, H.-L. A Novel H1N1 Virus Causes the First Pandemic of the 21st Century. *Eur. J. Immunol.* **2009**, *39*, 2946–2954. https://doi.org/10.1002/eji.200939911.
17. Van Kerkhove, M.D.; Ryan, M.J.; Ghebreyesus, T.A. Preparing for "Disease X". *Science* **2021**, *374*, 377–377. https://doi.org/10.1126/science.abm7796.
18. Iserson, K. The next Pandemic: Prepare for "Disease X". *West. J. Emerg. Med.* **2020**, *21*, 756. https://doi.org/10.5811/westjem.2020.5.48215.
19. Tahir, M.J.; Sawal, I.; Essar, M.Y.; Jabbar, A.; Ullah, I.; Ahmed, A. Disease X: A Hidden but Inevitable Creeping Danger. *Infect. Control Hosp. Epidemiol.* **2022**, *43*, 1758–1759. https://doi.org/10.1017/ice.2021.342.
20. Simpson, S.; Kaufmann, M.C.; Glozman, V.; Chakrabarti, A. Disease X: Accelerating the Development of Medical Countermeasures for the next Pandemic. *Lancet Infect. Dis.* **2020**, *20*, e108–e115. https://doi.org/10.1016/s1473-3099(20)30123-7.
21. Simpson, S.; Chakrabarti, A.; Robinson, D.; Chirgwin, K.; Lumpkin, M. Navigating Facilitated Regulatory Pathways during a Disease X Pandemic. *NPJ Vaccines* **2020**, *5*, 101. https://doi.org/10.1038/s41541-020-00249-5.
22. Radanliev, P.; De Roure, D. Disease X Vaccine Production and Supply Chains: Risk Assessing Healthcare Systems Operating with Artificial Intelligence and Industry 4.0. *Health Technol.* **2023**, *13*, 11–15. https://doi.org/10.1007/s12553-022-00722-2.
23. Singh, R.; Sarsaiya, S.; Singh, T.A.; Singh, T.; Pandey, L.K.; Pandey, P.K.; Khare, N.; Sobin, F.; Sikarwar, R.; Gupta, M.K. Corona Virus (COVID-19) Symptoms Prevention and Treatment: A Short Review. *J. Drug Deliv. Ther.* **2021**, *11*, 118–120. https://doi.org/10.22270/jddt.v11i2-s.4644.
24. Thakur, N. Sentiment Analysis and Text Analysis of the Public Discourse on Twitter about COVID-19 and MPox. *Big Data Cogn. Comput.* **2023**, *7*, 116. https://doi.org/10.3390/bdcc7020116.
25. Mercer, T.R.; Salit, M. Testing at Scale during the COVID-19 Pandemic. *Nat. Rev. Genet.* **2021**, *22*, 415–426. https://doi.org/10.1038/s41576-021-00360-w.
26. Kooli, C. COVID-19: Public Health Issues and Ethical Dilemmas. *Ethics Med. Public Health* **2021**, *17*, 100635. https://doi.org/10.1016/j.jemep.2021.100635.
27. Golan, M.S.; Jernegan, L.H.; Linkov, I. Trends and Applications of Resilience Analytics in Supply Chain Modeling: Systematic Literature Review in the Context of the COVID-19 Pandemic. *Environ. Syst. Decis.* **2020**, *40*, 222–243. https://doi.org/10.1007/s10669-020-09777-w.
28. Schuerger, C.; Batalis, S.; Quinn, K.; Adalja, A.; Puglisi, A. Viral Families and Disease X: A Framework for U.s. Pandemic Preparedness Policy. Available online: https://cset.georgetown.edu/wp-content/uploads/CSET-Viral-Families-and-Disease-X-A-Framework-for-U.S.-Pandemic-Preparedness-Policy.pdf (accessed on 17 August 2023).
29. Fontanet, A.; Cauchemez, S. COVID-19 Herd Immunity: Where Are We? *Nat. Rev. Immunol.* **2020**, *20*, 583–584. https://doi.org/10.1038/s41577-020-00451-5.
30. Frederiksen, L.S.F.; Zhang, Y.; Foged, C.; Thakur, A. The Long Road toward COVID-19 Herd Immunity: Vaccine Platform Technologies and Mass Immunization Strategies. *Front. Immunol.* **2020**, *11*, 1817. https://doi.org/10.3389/fimmu.2020.01817.
31. Kiviniemi, M.T.; Orom, H.; Hay, J.L.; Waters, E.A. Prevention Is Political: Political Party Affiliation Predicts Perceived Risk and Prevention Behaviors for COVID-19. *BMC Public Health* **2022**, *22*, 298. https://doi.org/10.1186/s12889-022-12649-4.
32. Rabin, C.; Dutra, S. Predicting Engagement in Behaviors to Reduce the Spread of COVID-19: The Roles of the Health Belief Model and Political Party Affiliation. *Psychol. Health Med.* **2022**, *27*, 379–388. https://doi.org/10.1080/13548506.2021.1921229.
33. Rasheed, R.T.; Mohammed, M.A.; Tapus, N. Big Data Analysis. *Mesopotamian J. Big Data* **2021**, *2021*, 22–25. https://doi.org/10.58496/mjbd/2021/004.
34. Yu, L.; Zhao, Y.; Tang, L.; Yang, Z. Online Big Data-Driven Oil Consumption Forecasting with Google Trends. *Int. J. Forecast.* **2019**, *35*, 213–223. https://doi.org/10.1016/j.ijforecast.2017.11.005.

35. Vaughan, L.; Chen, Y. Data Mining from Web Search Queries: A Comparison of Google Trends and Baidu Index: Data Mining from Web Search Queries: A Comparison of Google Trends and Baidu Index. *J. Assoc. Inf. Sci. Technol.* **2015**, *66*, 13–22. https://doi.org/10.1002/asi.23201.
36. Horák, J.; Ivan, I.; Kukuliač, P.; Inspektor, T.; Devečka, B.; Návratová, M. Google Trends for Data Mining. Study of Czech Towns. In *Computational Collective Intelligence. Technologies and Applications*; Springer: Berlin/Heidelberg, Germany, 2013; pp. 100–109; ISBN 9783642404948.
37. Sadeq, N.; Hamzeh, Z.; Nassreddine, G.; ElHassan, T. The Impact of Blockchain Technique on Trustworthy Healthcare Sector. *Mesopotamian J. Cyber Secur.* **2023**, *2023*, 105–115. https://doi.org/10.58496/mjcs/2023/015.
38. Tijerina, J.D.; Morrison, S.D.; Nolan, I.T.; Parham, M.J.; Richardson, M.T.; Nazerali, R. Celebrity Influence Affecting Public Interest in Plastic Surgery Procedures: Google Trends Analysis. *Aesthetic Plast. Surg.* **2019**, *43*, 1669–1680. https://doi.org/10.1007/s00266-019-01466-7.
39. Adawi, M.; Bragazzi, N.L.; Watad, A.; Sharif, K.; Amital, H.; Mahroum, N. Discrepancies between Classic and Digital Epidemiology in Searching for the Mayaro Virus: Preliminary Qualitative and Quantitative Analysis of Google Trends. *JMIR Public Health Surveill.* **2017**, *3*, e93. https://doi.org/10.2196/publichealth.9136.
40. Szmuda, T.; Ali, S.; Hetzger, T.V.; Rosvall, P.; Słoniewski, P. Are Online Searches for the Novel Coronavirus (COVID-19) Related to Media or Epidemiology? A Cross-Sectional Study. *Int. J. Infect. Dis.* **2020**, *97*, 386–390. https://doi.org/10.1016/j.ijid.2020.06.028.
41. Preis, T.; Moat, H.S.; Stanley, H.E. Quantifying Trading Behavior in Financial Markets Using Google Trends. *Sci. Rep.* **2013**, *3*, 1684. https://doi.org/10.1038/srep01684.
42. Kristoufek, L. BitCoin Meets Google Trends and Wikipedia: Quantifying the Relationship between Phenomena of the Internet Era. *Sci. Rep.* **2013**, *3*, 3415. https://doi.org/10.1038/srep03415.
43. Nghiem, L.T.P.; Papworth, S.K.; Lim, F.K.S.; Carrasco, L.R. Analysis of the Capacity of Google Trends to Measure Interest in Conservation Topics and the Role of Online News. *PLoS ONE* **2016**, *11*, e0152802. https://doi.org/10.1371/journal.pone.0152802.
44. Cho, S.; Sohn, C.H.; Jo, M.W.; Shin, S.-Y.; Lee, J.H.; Ryoo, S.M.; Kim, W.Y.; Seo, D.-W. Correlation between National Influenza Surveillance Data and Google Trends in South Korea. *PLoS ONE* **2013**, *8*, e81422. https://doi.org/10.1371/journal.pone.0081422.
45. Thakur, N.; Han, C.Y. Country-Specific Interests towards Fall Detection from 2004–2021: An Open Access Dataset and Research Questions. *Data* **2021**, *6*, 92. https://doi.org/10.3390/data6080092.
46. Miraz, M.H.; Ali, M.; Excell, P.S.; Picking, R. A Review on Internet of Things (IoT), Internet of Everything (IoE) and Internet of Nano Things (IoNT). In Proceedings of the 2015 Internet Technologies and Applications (ITA), Wrexham, UK, 8–11 September 2015; p. 219.
47. Adomavicius, G.; Tuzhilin, A. Toward the next Generation of Recommender Systems: A Survey of the State-of-the-Art and Possible Extensions. *IEEE Trans. Knowl. Data Eng.* **2005**, *17*, 734–749.
48. Jin, R.; Si, L.; Zhai, C. A Study of Mixture Models for Collaborative Filtering. *Inf. Retr. Boston.* **2006**, *9*, 357–382. https://doi.org/10.1007/s10791-006-4651-1.
49. Belk, M.; Papatheocharous, E.; Germanakos, P.; Samaras, G. Modeling Users on the World Wide Web Based on Cognitive Factors, Navigation Behavior and Clustering Techniques. *J. Syst. Softw.* **2013**, *86*, 2995–3012. https://doi.org/10.1016/j.jss.2013.04.029.
50. Eirinaki, M.; Vazirgiannis, M. Web Mining for Web Personalization. *ACM Trans. Internet Technol.* **2003**, *3*, 1–27. https://doi.org/10.1145/643477.643478.
51. Jalan, A.; Matkovskyy, R.; Urquhart, A.; Yarovaya, L. The Role of Interpersonal Trust in Cryptocurrency Adoption. *J. Int. Financ. Mark. Inst. Money* **2023**, *83*, 101715. https://doi.org/10.1016/j.intfin.2022.101715.
52. Mair, P.; Treiblmaier, H.; Lowry, P.B. Using Multistage Competing Risks Approaches to Model Web Page Transitions. *Internet Res.* **2017**, *27*, 650–669. https://doi.org/10.1108/intr-06-2016-0167.
53. Li, Y.; Zhu, T.; Li, A.; Zhang, F.; Xu, X. Web Behavior and Personality: A Review. In Proceedings of the 2011 3rd Symposium on Web Society, Port Elizabeth, South Africa, 26–28 October 2011.
54. Thakur, N.; Hall, I.; Han, C.Y. A Comprehensive Study to Analyze Trends in Web Search Interests Related to Fall Detection before and after COVID-19. In Proceedings of the 2022 5th International Conference on Computer Science and Software Engineering (CSSE 2022), New York, NY, USA, 21–23 October 2022.
55. Cervellin, G.; Comelli, I.; Lippi, G. Is Google Trends a Reliable Tool for Digital Epidemiology? Insights from Different Clinical Settings. *J. Epidemiol. Glob. Health* **2017**, *7*, 185. https://doi.org/10.1016/j.jegh.2017.06.001.
56. Teng, Y.; Bi, D.; Xie, G.; Jin, Y.; Huang, Y.; Lin, B.; An, X.; Feng, D.; Tong, Y. Dynamic Forecasting of Zika Epidemics Using Google Trends. *PLoS ONE* **2017**, *12*, e0165085. https://doi.org/10.1371/journal.pone.0165085.
57. Jun, S.-P.; Yoo, H.S.; Choi, S. Ten Years of Research Change Using Google Trends: From the Perspective of Big Data Utilizations and Applications. *Technol. Forecast. Soc. Change* **2018**, *130*, 69–87. https://doi.org/10.1016/j.techfore.2017.11.009.
58. Lippi, G.; Mattiuzzi, C.; Cervellin, G.; Favaloro, E.J. Direct Oral Anticoagulants: Analysis of Worldwide Use and Popularity Using Google Trends. Ann. Transl. Med. 2017, 5, 322–322, doi:10.21037/atm.2017.06.65.
59. Quintanilha, L.F.; Souza, L.N.; Sanchez, D.; Demarco, R.S.; Fukutani, K.F. The Impact of Cancer Campaigns in Brazil: A Google Trends Analysis. Ecancermedicalscience 2019, 13, doi:10.3332/ecancer.2019.963.
60. Nuti, S.V.; Wayda, B.; Ranasinghe, I.; Wang, S.; Dreyer, R.P.; Chen, S.I.; Murugiah, K. The Use of Google Trends in Health Care Research: A Systematic Review. *PLoS ONE* **2014**, *9*, e109583. https://doi.org/10.1371/journal.pone.0109583.
61. Dreher, P.C.; Tong, C.; Ghiraldi, E.; Friedlander, J.I. Use of Google Trends to Track Online Behavior and Interest in Kidney Stone Surgery. Urology 2018, 121, 74–78, doi:10.1016/j.urology.2018.05.040.

62. Ginsberg, J.; Mohebbi, M.H.; Patel, R.S.; Brammer, L.; Smolinski, M.S.; Brilliant, L. Detecting Influenza Epidemics Using Search Engine Query Data. *Nature* **2009**, *457*, 1012–1014. https://doi.org/10.1038/nature07634.
63. Kapitány-Fövény, M.; Ferenci, T.; Sulyok, Z.; Kegele, J.; Richter, H.; Vályi-Nagy, I.; Sulyok, M. Can Google Trends Data Improve Forecasting of Lyme Disease Incidence? *Zoonoses Public Health* **2019**, *66*, 101–107. https://doi.org/10.1111/zph.12539.
64. Verma, M.; Kishore, K.; Kumar, M.; Sondh, A.R.; Aggarwal, G.; Kathirvel, S. Google Search Trends Predicting Disease Outbreaks: An Analysis from India. *Healthc. Inform. Res.* **2018**, *24*, 300. https://doi.org/10.4258/hir.2018.24.4.300.
65. Young, S.D.; Torrone, E.A.; Urata, J.; Aral, S.O. Using Search Engine Data as a Tool to Predict Syphilis. *Epidemiology* **2018**, *29*, 574–578. https://doi.org/10.1097/ede.0000000000000836.
66. Young, S.D.; Zhang, Q. Using Search Engine Big Data for Predicting New HIV Diagnoses. *PLoS ONE* **2018**, *13*, e0199527. https://doi.org/10.1371/journal.pone.0199527.
67. Morsy, S.; Dang, T.N.; Kamel, M.G.; Zayan, A.H.; Makram, O.M.; Elhady, M.; Hirayama, K.; Huy, N.T. Prediction of Zika-Confirmed Cases in Brazil and Colombia Using Google Trends. *Epidemiol. Infect.* **2018**, *146*, 1625–1627. https://doi.org/10.1017/s0950268818002078.
68. Ortiz-Martínez, Y.; Garcia-Robledo, J.E.; Vásquez-Castañeda, D.L.; Bonilla-Aldana, D.K.; Rodriguez-Morales, A.J. Can Google® Trends Predict COVID-19 Incidence and Help Preparedness? The Situation in Colombia. *Travel Med. Infect. Dis.* **2020**, *37*, 101703. https://doi.org/10.1016/j.tmaid.2020.101703.
69. Vasconcellos-Silva, P.R.; Carvalho, D.B.F.; Trajano, V.; de La Rocque, L.R.; Sawada, A.C.M.B.; Juvanhol, L.L. Using Google Trends Data to Study Public Interest in Breast Cancer Screening in Brazil: Why Not a Pink February? *JMIR Public Health Surveill.* **2017**, *3*, e17. https://doi.org/10.2196/publichealth.7015.
70. Bragazzi, N.L.; Barberis, I.; Rosselli, R.; Gianfredi, V.; Nucci, D.; Moretti, M.; Salvatori, T.; Martucci, G.; Martini, M. How Often People Google for Vaccination: Qualitative and Quantitative Insights from a Systematic Search of the Web-Based Activities Using Google Trends. *Hum. Vaccin. Immunother.* **2017**, *13*, 464–469. https://doi.org/10.1080/21645515.2017.1264742.
71. Tkachenko, N.; Chotvijit, S.; Gupta, N.; Bradley, E.; Gilks, C.; Guo, W.; Crosby, H.; Shore, E.; Thiarai, M.; Procter, R.; et al. Google Trends Can Improve Surveillance of Type 2 Diabetes. *Sci. Rep.* **2017**, *7*, 4993. https://doi.org/10.1038/s41598-017-05091-9.
72. Carrière-Swallow, Y.; Labbé, F. Nowcasting with Google Trends in an Emerging Market: Nowcasting with Google Trends in an Emerging Market. *J. Forecast.* **2013**, *32*, 289–298. https://doi.org/10.1002/for.1252.
73. Thakur, N.; Han, C.Y. A Human-Human Interaction-Driven Framework to Address Societal Issues. In *Human Interaction, Emerging Technologies and Future Systems V*; Springer International Publishing: Cham, Switzerland, 2022; pp. 563–571; ISBN 9783030855390.
74. Thakur, N.; Han, C.Y. Google Trends to Investigate the Degree of Global Interest Related to Indoor Location Detection. In *Human Interaction, Emerging Technologies and Future Systems V*; Springer International Publishing: Cham, Switzerland, 2022; pp. 580–588; ISBN 9783030855390.
75. Kao, Y.-S. Do People Use ChatGPT to Replace Doctor? A Google Trends Analysis. *Ann. Biomed. Eng.* **2023**. https://doi.org/10.1007/s10439-023-03285-z.
76. Aslanidis, N.; Bariviera, A.F.; López, Ó.G. The Link between Cryptocurrencies and Google Trends Attention. *Fin. Res. Lett.* **2022**, *47*, 102654. https://doi.org/10.1016/j.frl.2021.102654.
77. Arezooji, D.M. A Big Data Analysis of the Ethereum Network: From Blockchain to Google Trends. *arXiv* **2021**, arXiv:2104.01764.
78. Padhi, S.S.; Pati, R.K. Quantifying Potential Tourist Behavior in Choice of Destination Using Google Trends. *Tour. Manag. Perspect.* **2017**, *24*, 34–47. https://doi.org/10.1016/j.tmp.2017.07.001.
79. Thakur, N.; Han, C.Y. An Intelligent Ubiquitous Activity Aware Framework for Smart Home. In *Human Interaction, Emerging Technologies and Future Applications III*; Springer International Publishing: Cham, Switzerland, 2021; pp. 296–302; ISBN 9783030553067.
80. Tran, U.S.; Andel, R.; Niederkrotenthaler, T.; Till, B.; Ajdacic-Gross, V.; Voracek, M. Low Validity of Google Trends for Behavioral Forecasting of National Suicide Rates. *PLoS ONE* **2017**, *12*, e0183149. https://doi.org/10.1371/journal.pone.0183149.
81. Thakur, N.; Han, C.Y. Indoor Localization for Personalized Ambient Assisted Living of Multiple Users in Multi-Floor Smart Environments. *Big Data Cogn. Comput.* **2021**, *5*, 42. https://doi.org/10.3390/bdcc5030042.
82. Sampri, A.; Mavragani, A.; Tsagarakis, K.P. Evaluating Google Trends as a Tool for Integrating the 'Smart Health' Concept in the Smart Cities' Governance in USA. *Procedia Eng.* **2016**, *162*, 585–592. https://doi.org/10.1016/j.proeng.2016.11.104.
83. Thakur, N.; Han, C.Y. Pervasive Activity Logging for Indoor Localization in Smart Homes. In Proceedings of the 2021 4th International Conference on Data Science and Information Technology, Shanghai, China, 23–25 July 2021.
84. Li, Y. How Is Data Visualization Shaping Our Life? The Application of Analytics from Google Trends during the Epidemic of COVID-19. In *Studies in Systems, Decision and Control*; Springer International Publishing: Cham, Switzerland, 2021; pp. 223–239; ISBN 9783030766313.
85. Thakur, N.; Han, C.Y. An Approach for Detection of Walking Related Falls during Activities of Daily Living. In Proceedings of the 2020 International Conference on Big Data, Artificial Intelligence and Internet of Things Engineering (ICBAIE), Fuzhou, China, 12–14 June 2020.
86. Kupfer, A.; Puhr, H. The Russian View on the War in Ukraine: Insights from Google Trends. *SSRN Electron. J.* **2022**. https://doi.org/10.2139/ssrn.4063194.
87. Artyukhov, A.; Barvinok, V.; Rehak, R.; Matvieieva, Y.; Lyeonov, S. Dynamics of Interest in Higher Education before and during Ongoing War: Google Trends Analysis. *Knowl. Perform. Manag.* **2023**, *7*, 47–63. https://doi.org/10.21511/kpm.07(1).2023.04.

88. Dolkar, T.; Gowda, S.; Chatterjee, S. Cardiac Symptoms during the Russia-Ukraine War: A Google Trends Analysis. *Cureus* **2023**, *15*, e36676. https://doi.org/10.7759/cureus.36676.
89. Faugère, C.; Gergaud, O. Business Ethics Searches: A Socioeconomic and Demographic Analysis of U.S. Google Trends in the Context of the 2008 Financial Crisis: Faugere and Gergaud. *Bus. Ethics* **2017**, *26*, 271–287. https://doi.org/10.1111/beer.12138.
90. Gao, J.; Xing, D.; Li, J.; Li, T.; Huang, C.; Wang, W. Is Robotic Assistance More Eye-Catching than Computer Navigation in Joint Arthroplasty? A Google Trends Analysis from the Point of Public Interest. *J. Robot. Surg.* **2023**, *17*, 2167–2176. https://doi.org/10.1007/s11701-023-01630-x.
91. Thakur, N.; Han, C.Y. A Multimodal Approach for Early Detection of Cognitive Impairment from Tweets. In *Human Interaction, Emerging Technologies and Future Systems V*; Springer International Publishing: Cham, Switzerland, 2022; pp. 11–19; ISBN 9783030855390.
92. Krishnan, N.; Anand, S.; Sandlas, G. Evaluating the Impact of COVID-19 Pandemic on Public Interest in Minimally Invasive Surgery: An Infodemiology Study Using Google Trends. *Cureus* **2021**, *13*, e18848. https://doi.org/10.7759/cureus.18848.
93. Thakur, N.; Han, C.Y. A Framework for Prediction of Cramps during Activities of Daily Living in Elderly. In Proceedings of the 2020 International Conference on Big Data, Artificial Intelligence and Internet of Things Engineering (ICBAIE), Fuzhou, China, 12–14 June 2020.
94. Google Trends. Available online: https://trends.google.com/trends/ (accessed on 18 August 2023).
95. Mavragani, A.; Ochoa, G. Google Trends in Infodemiology and Infoveillance: Methodology Framework. *JMIR Public Health Surveill.* **2019**, *5*, e13439. https://doi.org/10.2196/13439.
96. Arora, V.S.; McKee, M.; Stuckler, D. Google Trends: Opportunities and Limitations in Health and Health Policy Research. *Health Policy* **2019**, *123*, 338–341. https://doi.org/10.1016/j.healthpol.2019.01.001.
97. Mulero, R.; García-Hiernaux, A. Forecasting Spanish Unemployment with Google Trends and Dimension Reduction Techniques. *SERIEs* **2021**, *12*, 329–349. https://doi.org/10.1007/s13209-021-00231-x.
98. IEEE DataPort. Available online: https://ieee-dataport.org/ (accessed on 18 August 2023).
99. Wilkinson, M.D.; Dumontier, M.; Aalbersberg, I.J.; Appleton, G.; Axton, M.; Baak, A.; Blomberg, N.; Boiten, J.-W.; da Silva Santos, L.B.; Bourne, P.E.; et al. The FAIR Guiding Principles for Scientific Data Management and Stewardship. *Sci. Data* **2016**, *3*, 160018. https://doi.org/10.1038/sdata.2016.18.
100. Wishart, D.S.; Guo, A.; Oler, E.; Wang, F.; Anjum, A.; Peters, H.; Dizon, R.; Sayeeda, Z.; Tian, S.; Lee, B.L.; et al. HMDB 5.0: The Human Metabolome Database for 2022. *Nucleic Acids Res.* **2022**, *50*, D622–D631. https://doi.org/10.1093/nar/gkab1062.
101. Slenter, D.N.; Kutmon, M.; Hanspers, K.; Riutta, A.; Windsor, J.; Nunes, N.; Mélius, J.; Cirillo, E.; Coort, S.L.; Digles, D.; et al. WikiPathways: A Multifaceted Pathway Database Bridging Metabolomics to Other Omics Research. *Nucleic Acids Res.* **2018**, *46*, D661–D667. https://doi.org/10.1093/nar/gkx1064.
102. Banda, J.M.; Tekumalla, R.; Wang, G.; Yu, J.; Liu, T.; Ding, Y.; Artemova, E.; Tutubalina, E.; Chowell, G. A Large-Scale COVID-19 Twitter Chatter Dataset for Open Scientific Research—An International Collaboration. *Epidemiologia* **2021**, *2*, 315–324. https://doi.org/10.3390/epidemiologia2030024.
103. Thakur, N. A Large-Scale Dataset of Twitter Chatter about Online Learning during the Current COVID-19 Omicron Wave. *Data* **2022**, *7*, 109. https://doi.org/10.3390/data7080109.
104. Thakur, N. MonkeyPox2022Tweets: A Large-Scale Twitter Dataset on the 2022 Monkeypox Outbreak, Findings from Analysis of Tweets, and Open Research Questions. *Infect. Dis. Rep.* **2022**, *14*, 855–883. https://doi.org/10.3390/idr14060087.
105. Gjerding, M.N.; Taghizadeh, A.; Rasmussen, A.; Ali, S.; Bertoldo, F.; Deilmann, T.; Knøsgaard, N.R.; Kruse, M.; Larsen, A.H.; Manti, S.; et al. Recent Progress of the Computational 2D Materials Database (C2DB). *2D Mater.* **2021**, *8*, 044002. https://doi.org/10.1088/2053-1583/ac1059.
106. Kearnes, S.M.; Maser, M.R.; Wleklinski, M.; Kast, A.; Doyle, A.G.; Dreher, S.D.; Hawkins, J.M.; Jensen, K.F.; Coley, C.W. The Open Reaction Database. *J. Am. Chem. Soc.* **2021**, *143*, 18820–18826. https://doi.org/10.1021/jacs.1c09820.
107. Goodsell, D.S.; Zardecki, C.; Di Costanzo, L.; Duarte, J.M.; Hudson, B.P.; Persikova, I.; Segura, J.; Shao, C.; Voigt, M.; Westbrook, J.D.; et al. RCSB Protein Data Bank: Enabling Biomedical Research and Drug Discovery. *Protein Sci.* **2020**, *29*, 52–65. https://doi.org/10.1002/pro.3730.
108. Urban, M.; Cuzick, A.; Seager, J.; Wood, V.; Rutherford, K.; Venkatesh, S.Y.; De Silva, N.; Martinez, M.C.; Pedro, H.; Yates, A.D.; et al. PHI-Base: The Pathogen–Host Interactions Database. *Nucleic Acids Res.* **2019**, *48*, D613–D620. https://doi.org/10.1093/nar/gkz904.
109. Johnson, A.K.; Mehta, S.D. A Comparison of Internet Search Trends and Sexually Transmitted Infection Rates Using Google Trends. *Sex. Transm. Dis.* **2014**, *41*, 61–63. https://doi.org/10.1097/olq.0000000000000065.
110. Fazeli Dehkordy, S.; Carlos, R.C.; Hall, K.S.; Dalton, V.K. Novel Data Sources for Women's Health Research. *Acad. Radiol.* **2014**, *21*, 1172–1176. https://doi.org/10.1016/j.acra.2014.05.005.
111. Husnayain, A.; Fuad, A.; Lazuardi, L. Correlation between Google Trends on Dengue Fever and National Surveillance Report in Indonesia. *Glob. Health Action* **2019**, *12*, 1552652. https://doi.org/10.1080/16549716.2018.1552652.
112. Wang, D.; Guerra, A.; Wittke, F.; Lang, J.C.; Bakker, K.; Lee, A.W.; Finelli, L.; Chen, Y.-H. Real-Time Monitoring of Infectious Disease Outbreaks with a Combination of Google Trends Search Results and the Moving Epidemic Method: A Respiratory Syncytial Virus Case Study. *Trop. Med. Infect. Dis.* **2023**, *8*, 75. https://doi.org/10.3390/tropicalmed8020075.
113. Chan, E.H.; Sahai, V.; Conrad, C.; Brownstein, J.S. Using Web Search Query Data to Monitor Dengue Epidemics: A New Model for Neglected Tropical Disease Surveillance. *PLoS Negl. Trop. Dis.* **2011**, *5*, e1206. https://doi.org/10.1371/journal.pntd.0001206.

114. Alexander, D.J. Summary of Avian Influenza Activity in Europe, Asia, Africa, and Australasia, 2002–2006. *Avian Dis.* **2007**, *51*, 161–166. https://doi.org/10.1637/7602-041306r.1.
115. Bento, A.I.; Nguyen, T.; Wing, C.; Lozano-Rojas, F.; Ahn, Y.-Y.; Simon, K. Evidence from Internet Search Data Shows Information-Seeking Responses to News of Local COVID-19 Cases. *Proc. Natl. Acad. Sci. USA* **2020**, *117*, 11220–11222. https://doi.org/10.1073/pnas.2005335117.
116. Thakur, N.; Han, C. An Exploratory Study of Tweets about the SARS-CoV-2 Omicron Variant: Insights from Sentiment Analysis, Language Interpretation, Source Tracking, Type Classification, and Embedded URL Detection. *COVID* **2022**, *2*, 1026–1049. https://doi.org/10.3390/covid2080076.
117. Abeynayake, A.D.L.; Sunethra, A.A.; Deshani, K.A.D. A Stylometric Approach for Reliable News Detection Using Machine Learning Methods. In Proceedings of the 2022 22nd International Conference on Advances in ICT for Emerging Regions (ICTer), Colombo, Sri Lanka, 30 November–1 December 2022.
118. Health Ministry Says on High Alert for Any Possible Existence of 'Disease X' in Malaysia. Available online: https://www.theborneopost.com/2023/09/30/health-ministry-says-on-high-alert-for-any-possible-existence-of-disease-x-in-malaysia/ (accessed on 3 October 2023).
119. Explainers, F.P. "Deadlier than COVID": How Dangerous Is Disease X? Available online: https://www.firstpost.com/explainers/deadlier-than-covid-how-dangerous-is-disease-x-13192892.html (accessed on 3 October 2023).
120. Live "Disease X" Could Be 20 Times Deadlier than COVID-19, Says Expert. Top 10 Updates. Available online: https://www.livemint.com/science/health/disease-x-could-be-20-times-deadlier-than-covid-19-says-expert-top-10-updates-11695606507951.html (accessed on 3 October 2023).
121. Vats, V. What Is Disease X? It Could Bring the next Pandemic, Says Expert. Available online: https://www.ndtv.com/health/what-is-disease-x-it-could-bring-the-next-pandemic-deadlier-than-covid-19-says-expert-4424840 (accessed on 3 October 2023).